\documentclass[11pt,a4paper]{article}
\pdfoutput=1
\usepackage{jheppub}
\usepackage[T1]{fontenc}
\usepackage{microtype}
\usepackage{amsmath,amssymb,mathtools,bm}
\usepackage{booktabs,array,multirow}
\usepackage{graphicx}
\usepackage{tikz}
\usetikzlibrary{arrows.meta,angles,calc,quotes,decorations.pathmorphing,decorations.pathreplacing,backgrounds,shadings}
\usepackage{xcolor}
\definecolor{jhepNavy}{HTML}{1B365D}
\definecolor{jhepCrimson}{HTML}{B22222}
\definecolor{jhepTeal}{HTML}{007A78}
\definecolor{jhepAmber}{HTML}{D97706}
\definecolor{jhepPurple}{HTML}{6B4C9A}
\definecolor{jhepSlate}{HTML}{4A5568}
\graphicspath{{./}}

\newcommand{\dd}{\mathrm{d}}
\newcommand{\gev}{\,\mathrm{GeV}}
\newcommand{\tev}{\,\mathrm{TeV}}
\newcommand{\bk}{\bm{k}}
\newcommand{\bp}{\bm{p}}
\newcommand{\bq}{\bm{q}}
\newcommand{\bK}{\bm{K}}
\newcommand{\bP}{\bm{P}}
\newcommand{\bb}{\bm{b}}
\newcommand{\bl}{\bm{l}}
\newcommand{\br}{\bm{r}}
\newcommand{\Qs}{Q_s}
\newcommand{\CNP}{C_{\mathrm{NP}}}
\newcommand{\aT}{a_T}
\newcommand{\Kperp}{K_\perp}
\newcommand{\SigmaHF}{\Sigma_{\mathrm{HF}}}
\newcommand{\pPb}{p+\mathrm{Pb}}
\newcommand{\pO}{p+\mathrm{O}}
\newcommand{\pp}{p+p}
\newcommand{\DD}{D^0\bar{D}^0}
\newcommand{\BB}{B^0\bar{B}^0}

\title{Heavy-flavor energy correlators and transverse-momentum projections in forward proton-nucleus collisions}
\hypersetup{
  pdftitle={Heavy-flavor energy correlators and transverse-momentum projections in forward proton-nucleus collisions},
  pdfauthor={Lei Wang},
  pdfsubject={JHEP manuscript},
  pdfkeywords={gluon saturation, heavy flavor, energy correlators, Sudakov resummation}
}

\author[a]{Lei Wang}

\affiliation[a]{Key Laboratory of Quark and Lepton Physics (MOE)\\
and Institute of Particle Physics, Central China Normal University,\\
No.~152 Luoyu Road, Wuhan, Hubei 430079, China}

\emailAdd{leiwang@ccnu.edu.cn}

\abstract{%
Forward heavy-meson pair correlations in proton-nucleus collisions probe small-$x$ gluon saturation in the nuclear target, while their azimuthal distribution is shaped by soft-collinear radiation, heavy-quark fragmentation, and double parton scattering (DPS). We study heavy-flavor-tagged transverse energy--energy correlators (HF-TEEC), their mirror-antisymmetric combination (HF-ATEEC), and the orthogonal pair-imbalance projection $\Kperp \equiv \aT$ for forward $\DD$ and $\BB$ production. Transverse-energy weighting shifts the effective fragmentation moment toward larger momentum fractions and reduces the relative DPS background, while HF-ATEEC removes azimuthally flat and even components identically. Projecting the pair imbalance orthogonal to the di-meson axis eliminates linear longitudinal fragmentation recoil; in the factorized back-to-back limit, its one-dimensional Fourier-cosine transform also cancels common Sudakov factors. We evaluate these observables in the massive hybrid Color Glass Condensate framework with running-coupling Balitsky--Kovchegov dipoles, threshold-matched SCET/bHQET Sudakov resummation, and finite-mass fragmentation. The leading-order calculation reproduces the away-side shape of forward LHCb $\DD$ data in $\pp$ and $\pPb$ collisions. At $\sqrt{s_{NN}} = 8.16\tev$, we compare saturation predictions for $\pp$, $\pO$, and $\pPb$ collisions with collinear nuclear shadowing (\texttt{nNNPDF3.0}) and linear Gaussian multiple scattering. Nonlinear saturation produces stronger $\Kperp$ broadening than both baselines, with $\BB$ pairs showing the largest separation. This hierarchy persists under variations of the nonperturbative Sudakov parameter, matching scale, and detector resolution.
}


\begin{document}
\maketitle
\raggedbottom

\section{Introduction}
\label{sec:intro}

Forward particle production in proton-nucleus ($\pPb$) collisions at the Large Hadron Collider (LHC) probes a strongly asymmetric partonic configuration. The projectile parton from the proton carries a moderate or large longitudinal momentum fraction $x_p \sim 10^{-1}$ and is described by collinear parton distribution functions, whereas the nuclear target is probed at small $x_A \sim 10^{-5}\text{--}10^{-4}$. In this small-$x$ regime, the rapid growth of the gluon density driven by linear bremsstrahlung is tamed by nonlinear gluon recombination, generating a semi-hard saturation scale $\Qs(x_A) \gtrsim 1\gev$ described by the Color Glass Condensate (CGC) effective field theory~\cite{Kuraev:1977fs,Gribov:1983ivg,Mueller:1985wy,McLerran:1993ni,McLerran:1993ka,McLerran:1994vd,Balitsky:1995ub,Kovchegov:1999yj,Jalilian-Marian:1997jx,Jalilian-Marian:1997gr,Iancu:2000hn,Iancu:2001ad,Ferreiro:2001qy,Iancu:2003xm,Weigert:2005us,Jalilian-Marian:2005ccm,Balitsky:2006wa,Gelis:2010nm,Kovchegov:2012mbw,Albacete:2014fwa,Blaizot:2016qgz,Petreska:2018cbf,Morreale:2021pnn,Boussarie:2023izj,Garcia-Montero:2025bpn,Boussarie:2025rot,Iancu:2026cgc}. Over the past two decades, the dilute-dense hybrid factorization formalism and its next-to-leading-order (NLO) extensions have provided the primary theoretical framework for forward single-inclusive hadron and jet phenomenology~\cite{Dumitru:2002qt,Dumitru:2005gt,Albacete:2010bs,Altinoluk:2011qy,Chirilli:2011km,Chirilli:2012jd,Watanabe:2015yca,Iancu:2016vyg,Ducloue:2016shw,Liu:2020mpy,Iancu:2020mos,Shi:2021hwx,Liu:2022ijp,Wang:2022fwj,Duan:2025mult,Wang:2026mult}. Compared with single-inclusive yields, forward two-particle azimuthal correlations offer a more differential test of the target color field: in nearly back-to-back kinematics, the pair transverse-momentum imbalance is generated directly by coherent multiple gluon exchanges encoded in small-$x$ Wilson-line correlators and transverse-momentum-dependent (TMD) gluon distributions~\cite{Kharzeev:2004bw,Marquet:2007vb,Albacete:2010pg,Dominguez:2010xd,Dominguez:2011wm,Jalilian-Marian:2012wwi,Lappi:2012nh,Zheng:2014vka,Kotko:2015ura,Lappi:2015fma,Albacete:2018ruq,Stasto:2018rci,Altinoluk:2019fui,vanHameren:2019ysa,Fujii:2020bkl,Shi:2020collect,Caucal:2021ent,Taels:2022tza,Benic:2022ixp,Al-Mashad:2022zbq,Tong:2022zwp,Bergabo:2023wed,Deganutti:2023qjm,Wang:2026efm,Caucal:2026cs}.

In forward light-hadron and dijet measurements~\cite{STAR:2021fgw,Caucal:2025qsa}, however, isolating the nuclear saturation recoil from soft-collinear gluon radiation remains challenging. Whenever the pair transverse-momentum imbalance $|\bK|$ is much smaller than the hard invariant mass $M_{\mathrm{inv}}$ of the pair, large double and single Sudakov logarithms $\alpha_s^m \ln^{2m}(M_{\mathrm{inv}}^2/\bK^2)$ arise from soft and collinear emissions. A consistent calculation in the back-to-back region must therefore resum these Sudakov logarithms through the Collins--Soper--Sterman (CSS) formalism alongside small-$x$ Balitsky--Kovchegov (BK) or JIMWLK evolution~\cite{Collins:1984kg,Collins:2011zzd,Mueller:2013wwa,Sun:2014gfa,Chen:2016vem,Liu:2020mpy,Shi:2021hwx,Caucal:2022ulg,Wang:2022fwj,Caucal:2023nci}. For massless partons at hard scales $M_{\mathrm{inv}} \gtrsim 10\gev$, perturbative and nonperturbative Sudakov radiation substantially broadens the away-side peak in both $\pp$ and $\pPb$ collisions. Because this vacuum-like radiative recoil acts identically on proton and nuclear targets, it dilutes the relative broadening induced by the target saturation momentum $|\bK'| \sim \Qs(x_A)$. As a consequence, the nuclear modification observed in light-hadron angular correlations often reduces to an overall suppression of the away-side per-trigger yield, where nonlinear saturation competes with collinear nuclear parton distribution function (nPDF) shadowing~\cite{Kovarik:2015cma,Eskola:2021nhw,AbdulKhalek:2022fyi} and light-hadron fragmentation effects.

Open heavy-flavor meson pairs ($\DD$ and $\BB$) alter this scale hierarchy and provide a cleaner window on the target transverse-momentum distribution. At forward rapidity at the LHC, heavy-quark pair production in perturbative QCD~\cite{Nason:1987xz,Nason:1989zy,Cacciari:1998it,Cacciari:2001td,Cacciari:2012ny,Helenius:2018uul,Helenius:2023wkn} and in the CGC~\cite{Blaizot:2004wv,Fujii:2006ab,Kovchegov:2006qn,Fujii:2013yja,Altinoluk:2015vax,Marquet:2017xwy,Ma:2018bax,Zhang:2020collect,delCastillo:2020omr,Altinoluk:2020qic} is governed by the gluon splitting channel $g + A \to Q\bar{Q} + X$. Recent theoretical developments~\cite{Marquet:2025msx,Gao:2026msx,Salazar:2026ssz} have combined this massive hybrid-CGC cross section with Soft-Collinear Effective Theory and boosted Heavy Quark Effective Theory (SCET/bHQET) Sudakov resummation~\cite{Bauer:2000ew,Bauer:2000yr,Bauer:2001yt,Fleming:2007qr,Fleming:2007xt,Becher:2010tm,Hoang:2015vua,vonKuk:2023jfd,Dai:2026dls} and finite-mass heavy-meson fragmentation~\cite{Peterson:1982ak,Cacciari:2005uk,Kneesch:2007ey,Kniehl:2008zza,Kniehl:2012ti}. Compared with light hadrons, heavy-flavor pairs offer three physical advantages. First, the finite heavy-quark mass $m_Q$ screens collinear gluon radiation inside the angular dead cone $\theta \lesssim m_Q / E$~\cite{Dokshitzer:1991fd,ALICE:2021aqk}, freezing the final-state cusp double logarithm below $\bar{\mu} = m_Q$ and reducing perturbative Sudakov broadening for bottom quarks. Second, heavy-quark fragmentation functions peak at large momentum fractions ($\langle\bar{\xi}\rangle \simeq 0.66$ for $c \to D^0$ and $\langle\bar{\xi}\rangle \simeq 0.80\text{--}0.81$ for $b \to B^0$), so the reconstructed meson pair preserves the transverse recoil of the parent $Q\bar{Q}$ pair much more faithfully than light pions. Third, comparing charm ($m_c = 1.28\gev$) and bottom ($m_b = 4.50\gev$) pairs within the same detector acceptance provides a built-in mass lever arm that tests both the hard production kernel and the scale dependence of nuclear broadening~\cite{Salazar:2026ssz}.

A parallel line of research has reformulated angular correlations in terms of infrared- and collinear-safe energy-flow observables. Energy--energy correlators (EEC) in $e^+e^-$ annihilation~\cite{Basham:1978bw,Basham:1978zq} and transverse energy--energy correlators (TEEC) in hadron collisions~\cite{Ali:1984yp,ATLAS:2015yaa,ATLAS:2023tgo} weight pair angular distributions by products of particle energies or transverse energies. In recent years, energy correlators have been widely applied to perturbative jet substructure, heavy-quark dead-cone radiation, and high-logarithmic resummation~\cite{Moult:2018jzp,Gao:2019ojf,Dixon:2019uzg,Chen:2020vvp,Li:2020bub,Ebert:2020sjp,Craft:2022kdo,Gao:2023ivm,CMS:2024mlf,ALICE:2024dfl,Liu:2024eec,STAR:2025jut,ALICE:2025eec,Chang:2025eec,Guo:2025zto,Kang:2025zto,Barata:2025hqe,Song:2025eec,Duan:2026eec,Barata:2026ztg,Jobilal:2026eec}, as well as to nucleon energy correlators~\cite{Liu:2023aqb,Liu:2023onb,Mantysaari:2026mstt}, cold-nuclear-matter and heavy-ion jet modifications~\cite{Andres:2022ovj,Yang:2023dwc,Barata:2023bhh,Shen:2024eec,Barata:2024wsu,Fu:2024eec,CMS:2025eec,Ke:2025eec,Barata:2026bls}, and TEECs in deep inelastic scattering and $\pPb$ collisions~\cite{Kang:2024klsz,Kang:2024kpz,Kang:2025klp,Kang:2025kklp,Bhattacharya:2025bkp,Ganguli:2025gk,Shao:2026ssz}. In addition, single-axis transverse-momentum projections ($\aT$), one-dimensional small-acoplanarity recoil factorizations, and two-scale ratios have been developed to isolate transverse recoil from longitudinal smearing and Sudakov dilution~\cite{Banfi:2010xy,Lin:2026lz,Ke:2026kll,Jobilal:2026eec}. Very recently, ref.~\cite{Kang:2026klp} evaluated the inclusive all-hadron TEEC for forward heavy-quark pair production in the CGC at leading order without Sudakov resummation; by summing over all final-state hadron species and integrating over all hadron transverse momenta, that formulation uses the collinear momentum sum rule to eliminate fragmentation functions in the intermediate angular region $0.1 \le \tau \le 0.9$. Forward LHCb measurements, however, reconstruct identified $\DD$ and $\BB$ pairs within finite transverse-momentum windows~\cite{LHCb:2012aiv,LHCb:2013vjr,LHCb:2016ikn,LHCb:2017yua,LHCb:2017bmm,LHCb:2019avm,LHCb:2020jse}. In such flavor-tagged measurements, the collinear sum rule does not apply, and the primary sensitivity to the nuclear saturation scale $\Qs(x_A)$ is concentrated near the back-to-back endpoint $\tau \to 0$, where massive Sudakov resummation, finite-mass fragmentation, and multiparton backgrounds enter together.

Extracting a clean small-$x$ saturation signal from identified heavy-meson pairs requires addressing three physical sources of background that compete with coherent nuclear recoil. First, independent fluctuations of the two heavy-quark fragmentation fractions ($\xi_1 \neq \xi_2$) generate a longitudinal recoil along the di-meson axis at linear order in the fragmentation asymmetry $\epsilon = (\xi_1 - \xi_2)/(\xi_1 + \xi_2)$, even when the parent $Q\bar{Q}$ pair is back-to-back ($\Delta\Phi = \pi$). This longitudinal recoil shifts the total vector imbalance $|\bq_T| = |\bp_1 + \bp_2|$ away from zero and masks the transverse kick $|\bK'|$ from the nuclear target. Second, uncorrelated Double Parton Scattering (DPS) is enhanced in $\pPb$ relative to $\pp$ collisions by a factor $F_{\mathrm{DPS}}^{p\mathrm{Pb}} \simeq 3\text{--}4$ because the two hard gluon-fusion processes can occur on different target nucleons along the projectile impact parameter~\cite{Strikman:2001gz,dEnterria:2012jam,Helenius:2019uge,Blok:2019fgg,LHCb:2020jse}. The resulting azimuthally flat DPS pedestal is prominent in LHCb $\pPb$ $\DD$ data around $\Delta\Phi \simeq \pi/2$ and biases away-side width extractions if left unsubtracted. Third, the nonperturbative CSS Sudakov exponent depends on a phenomenological parameter $\CNP \in [1.5, 3.5]$~\cite{Sun:2014dqm} that broadens individual $\pp$ and $\pPb$ spectra. A diagnostic observable for LHCb Run~3 and Run~4 must therefore suppress longitudinal fragmentation smearing, eliminate the DPS pedestal, and remain stable under variations of $\CNP$, while distinguishing nonlinear rcBK saturation from collinear nPDF shadowing and linear Gaussian multiple scattering.

In this work, we develop a unified analysis framework based on the heavy-flavor-tagged TEEC (HF-TEEC), its mirror-antisymmetric combination (HF-ATEEC), and the orthogonal pair-imbalance projection $\Kperp \equiv \aT$ for forward $\DD$ and $\BB$ production. Each component targets a specific background mechanism. Weighting the tagged pair cross section by $(E_{T1}E_{T2})^n$ shifts the sampled fragmentation moment toward larger $\xi$ and suppresses the relative DPS pedestal across wide $p_T$ intervals, while subtracting the mirrored near-side hemisphere in HF-ATEEC cancels any flat DPS pedestal and all even azimuthal harmonics algebraically. In transverse-momentum space, projecting the pair imbalance $\bq_T$ onto the axis orthogonal to the di-meson direction $\bP_h = \frac{1}{2}(\bp_1 - \bp_2)$ removes linear $\mathcal{O}(\epsilon)$ longitudinal fragmentation recoil exactly, yielding a distribution that peaks smoothly at $\Kperp = 0$. Furthermore, in the factorized back-to-back limit, the one-dimensional Fourier-cosine transform of the $\Kperp$ distribution reduces the two-dimensional impact-parameter Hankel transform to a single cosine integral and cancels common perturbative and nonperturbative Sudakov factors in the $\pPb/\pp$ ratio. We implement these observables in the massive hybrid-CGC framework using HERA-constrained running-coupling BK dipoles, threshold-matched SCET/bHQET massive Sudakov resummation, and finite-mass heavy-meson fragmentation. After benchmarking the leading-order angular distributions and HF-ATEEC against LHCb forward $\DD$ measurements in $\pp$ ($\sqrt{s}=7\tev$) and $\pPb$ ($\sqrt{s_{NN}}=8.16\tev$) collisions, we present systematic predictions at $\sqrt{s_{NN}} = 8.16\tev$ across $\pp$, light-ion $\pO$, and $\pPb$ collisions and contrast them against collinear \texttt{nNNPDF3.0} shadowing and linear Gaussian broadening.

The remainder of this paper is organized as follows. Section~\ref{sec:framework} reviews the finite-mass forward $g+A \to Q\bar{Q}+X$ kinematics, the large-$N_c$ hybrid-CGC cross section with threshold-matched SCET/bHQET Sudakov resummation, and the construction of the rcBK, \texttt{nNNPDF3.0}, and Gaussian-broadening target baselines. Section~\ref{sec:teec} defines the order-$n$ HF-TEEC and mirror-antisymmetric HF-ATEEC, and derives their Mellin-moment fragmentation-hardening and DPS-cancellation identities. Section~\ref{sec:proj} introduces the orthogonal projection $\Kperp \equiv \aT$, proves the cancellation of linear fragmentation recoil, and formulates the one-dimensional Fourier-cosine transform and two-scale double ratio. Section~\ref{sec:results} presents our numerical results as a connected validation chain: we begin with the LHCb $\DD$ data benchmarks (section~\ref{sec:lhcb_bench}), proceed to the $\sqrt{s_{NN}}=8.16\tev$ HF-TEEC ratios and smooth $\Kperp$ spectra across flavor and $p_T$ windows (section~\ref{sec:results_kperp}), isolate the physical origin of the charm--bottom hierarchy through a dead-cone mechanism decomposition (section~\ref{sec:results_deadcone}), examine the finite-acceptance cosine transform and two-scale double ratio (section~\ref{sec:results_cosine}), and conclude with the $\pp \to \pO \to \pPb$ nuclear-opacity scan and systematic uncertainty analysis (section~\ref{sec:results_sys}). Section~\ref{sec:conclusions} summarizes our main findings and outlines future extensions. Appendix~\ref{app:kperp} details the analytic $\Kperp$ inversion, the exact change-of-variables Jacobian, and the numerical verification suite.

\section{Hybrid CGC framework, dead-cone Sudakov, and nuclear baselines}
\label{sec:framework}

To place the observables introduced in sections~\ref{sec:teec} and~\ref{sec:proj} on a consistent theoretical footing, we first specify the partonic production kernel, the soft-collinear radiation treatment, and the target models used throughout our numerical study. Because forward heavy-meson pairs at LHCb are measured at transverse momenta $p_T \sim 2\text{--}8\gev$ comparable to the charm and bottom masses, mass effects enter simultaneously in three places: the parton-to-meson fragmentation kinematics, the hard gluon-splitting matrix element, and the dead-cone threshold in the Sudakov anomalous dimension. We summarize each ingredient below, followed by the nuclear geometry and non-saturation reference baselines.

\subsection{\texorpdfstring{Forward $g+A \to Q\bar{Q}+X$ kinematics and finite-mass fragmentation}{Forward g plus A to Q Qbar plus X kinematics and finite-mass fragmentation}}
\label{sec:kinematics}

We consider inclusive forward heavy-meson pair production in proton-proton and proton-nucleus collisions,
\begin{equation}
  p + A \longrightarrow H_1(\bp_1, y_1) + \bar{H}_2(\bp_2, y_2) + X,
  \qquad H \in \{D^0, B^0\},
  \label{eq:process}
\end{equation}
with both mesons reconstructed in the proton-going forward hemisphere. For our LHC Run~3--4 predictions we adopt the fiducial acceptance $2.7 < y_1, y_2 < 3.7$ at $\sqrt{s_{NN}} = 8.16\tev$, while for the experimental benchmarks in section~\ref{sec:lhcb_bench} we match the center-of-mass energies and fiducial cuts of the published LHCb $\DD$ measurements~\cite{LHCb:2012aiv,LHCb:2020jse}. Following ref.~\cite{Salazar:2026ssz}, we use heavy-quark and heavy-meson masses $m_c = 1.28\gev$, $M_D = 1.86484\gev$ for charm, and $m_b = 4.50\gev$, $M_B = 5.27966\gev$ for bottom.

At transverse momenta comparable to the heavy-meson mass $M_H$, the difference between the quark mass $m_Q$ and the meson mass $M_H$ modifies the longitudinal and transverse momentum sharing during hadronization. Let $E_{Ti} = \sqrt{p_{Ti}^2 + M_H^2}$ denote the meson transverse energy, $p_{zi} = E_{Ti}\sinh y_i$ its longitudinal momentum, and $p_i = \sqrt{p_{Ti}^2 + p_{zi}^2}$ its three-momentum magnitude in the nucleon-nucleon center-of-mass frame. We employ the finite-mass collinear fragmentation map of ref.~\cite{Salazar:2026ssz}, which preserves the three-momentum direction of the parent heavy quark while scaling the light-cone momentum along each particle's flight axis as $p_{i,\parallel}^+ \equiv E_i + |\vec{p}_i| = \xi_i k_{i,\parallel}^+$. For a reconstructed meson with kinematics $(p_{Ti}, y_i)$ and fragmentation fraction $\xi_i \in (0,1)$, the parent heavy-quark transverse momentum $\bk_i$ and rapidity $y_{qi}$ are fixed uniquely by
\begin{align}
  k_{i,\parallel}^+ &= \frac{E_{Ti}\cosh y_i + p_i}{\xi_i}, &
  \lambda_i &= \frac{k_{i,\parallel}^+ - m_Q^2 / k_{i,\parallel}^+}{2 p_i}, \nonumber\\
  k_{Ti} &= \lambda_i p_{Ti}, &
  y_{qi} &= \operatorname{arsinh}\!\left(\frac{\lambda_i p_{zi}}{\sqrt{k_{Ti}^2 + m_Q^2}}\right),
  \label{eq:frag_map}
\end{align}
with three-dimensional phase-space Jacobian $\mathcal{J}_i \equiv |\partial(\bk_i, y_{qi})/\partial(\bp_i, y_i)| = \lambda_i^2$~\cite{Salazar:2026ssz}. In the high-momentum limit $p_{Ti} \gg M_H, m_Q$, eq.~\eqref{eq:frag_map} reduces smoothly to the familiar massless collinear relations $\lambda_i \to 1/\xi_i$, $p_{Ti} \simeq \xi_i k_{Ti}$, and $y_{qi} \simeq y_i$, whereas at low $p_{Ti}$ it accounts for the threshold energy cost of producing a massive meson of mass $M_H > m_Q$.

Once the parent heavy-quark kinematics $(k_{Ti}, y_{qi})$ are determined by eq.~\eqref{eq:frag_map}, their transverse masses $m_{Ti} = \sqrt{k_{Ti}^2 + m_Q^2}$ fix the projectile and target longitudinal momentum fractions $x_p$ and $x_A$ as well as the longitudinal plus-momentum sharing fraction $z$:
\begin{equation}
  x_p = \frac{m_{T1} e^{y_{q1}} + m_{T2} e^{y_{q2}}}{\sqrt{s_{NN}}}, \qquad
  x_A = \frac{m_{T1} e^{-y_{q1}} + m_{T2} e^{-y_{q2}}}{\sqrt{s_{NN}}}, \qquad
  z = \frac{m_{T1} e^{y_{q1}}}{m_{T1} e^{y_{q1}} + m_{T2} e^{y_{q2}}},
  \label{eq:parton_kin}
\end{equation}
with hard scale $\mu = \sqrt{s_{NN} x_p x_A}$. In terms of the total partonic transverse momentum $\bK = \bk_1 + \bk_2$ and the relative partonic transverse momentum $\bP = (1-z)\bk_1 - z\bk_2$, the invariant mass squared of the $Q\bar{Q}$ pair satisfies the exact relation
\begin{equation}
  M_{\mathrm{inv}}^2 = \mu^2 - \bK^2 = \frac{\bP^2 + m_Q^2}{z(1-z)}.
  \label{eq:minv_exact}
\end{equation}
Using the right-hand side of eq.~\eqref{eq:minv_exact} in numerical evaluations guarantees positive-definite invariant masses and avoids subtractive cancellation when $|\bK|$ approaches $\mu$.

For the collinear projectile gluon distribution $g_p(x_p, \mu)$ and the strong running coupling $\alpha_s(\mu)$, we use the CT18NNLO set~\cite{Hou:2019efy}. Charm-quark fragmentation $c \to D^0$ is described by the next-to-leading-order KKKS08 parameterization~\cite{Kneesch:2007ey}. For bottom-quark fragmentation $b \to B^0$, we adopt the KKSS Kartvelishvili initial condition $D_{B/b}(\xi, \mu_0 = 4.5\gev) \propto \xi^{16.87}(1-\xi)^{2.628}$, normalized to the experimental branching fraction $f(b \to B^0) = 0.404$~\cite{Kniehl:2008zza}, and evolve it on a dense $(\xi, \mu)$ grid up to $\mu = 200\gev$ via fourth-order Runge--Kutta nonsinglet DGLAP evolution. Because $D_{B/b}(\xi, \mu)$ peaks at significantly larger momentum fractions than $D_{D/c}(\xi, \mu)$, bottom mesons carry a larger fraction of their parent quark momentum, which both reduces fragmentation smearing and lowers the parent parton momentum $k_T \simeq p_T/\xi$ required to populate a given hadron-$p_T$ bin.

\subsection{\texorpdfstring{Large-$N_c$ hybrid CGC cross section and threshold-matched Sudakov}{Large-Nc hybrid CGC cross section and threshold-matched Sudakov}}
\label{sec:cgc_xsec}

Having established the parton-to-meson kinematic mapping, we now connect the parent $Q\bar{Q}$ state to the small-$x$ target color field. In the dilute-dense hybrid CGC framework, forward heavy-quark pair production proceeds via a collinear gluon from the proton projectile splitting into a $Q\bar{Q}$ color dipole before or after eikonal scattering off the dense nuclear shockwave~\cite{Blaizot:2004wv,Fujii:2006ab,Kovchegov:2006qn,Fujii:2013yja,Altinoluk:2015vax}. In the large-$N_c$ approximation, the four-point quadrupole and double-dipole Wilson-line correlators entering the cross section factorize into a product of two fundamental dipole amplitudes~\cite{Marquet:2007vb,Dominguez:2011wm,Lappi:2015fma,Salazar:2026ssz}. Convoluting the partonic cross section with the finite-mass fragmentation functions and the Sudakov matching kernel yields the differential heavy-meson pair cross section
\begin{align}
  \frac{\dd\sigma^{pA \to H\bar{H}X}}{\dd^2\bp_1\,\dd y_1\,\dd^2\bp_2\,\dd y_2}
  \propto \int_0^1 \dd\xi_1\,\dd\xi_2\,\lambda_1^2\lambda_2^2\, z(1-z)\,\alpha_s(\mu)
  \int \dd^2\bl_1\,\dd^2\bl_2\,& F_A(x_A, \bl_1)\, F_A(x_A, \bl_2) \nonumber\\
  &\times \mathcal{H}_{gA}\, W_{\mathrm{match}}(|\bK - \bK'|),
  \label{eq:master_xsec}
\end{align}
where the momentum-fraction integrals are restricted to the physical domain $x_p, x_A < 1$, and $\bK' = \bl_1 + \bl_2$ represents the net transverse momentum transferred from the two target dipole scatterings to the $Q\bar{Q}$ pair. The momentum-space fundamental dipole distribution $F_A(x_A, \bl)$ is the two-dimensional Fourier transform of the coordinate-space dipole $S$-matrix $S_A(Y_A, r) = 1 - N_A(Y_A, r)$ evaluated at evolution rapidity $Y_A = \ln(0.01/x_A)$,
\begin{equation}
  F_A(x_A, \bl) = \int \frac{\dd^2\br}{(2\pi)^2}\, e^{-i\bl\cdot\br}\, S_A(Y_A, r),
  \qquad \int \dd^2\bl\, F_A(x_A, \bl) = S_A(Y_A, 0) = 1.
  \label{eq:dipole_ft}
\end{equation}
To evaluate the oscillatory Hankel transform over $l \in [0, 50]\gev$ without numerical ringing or ultraviolet aliasing, we interpolate $S_A(Y_A, r)$ with a $C^2$ cubic spline in $\ln r$, apply 12-point sub-interval Gauss--Legendre quadrature for $l \le 4\gev$, and switch to an analytic integration-by-parts curvature transform for $l > 4\gev$. This hybrid quadrature preserves the unit probability sum rule $|\int\dd^2\bl\,F_A - 1| < 5\times 10^{-4}$ across all proton and nuclear grids.

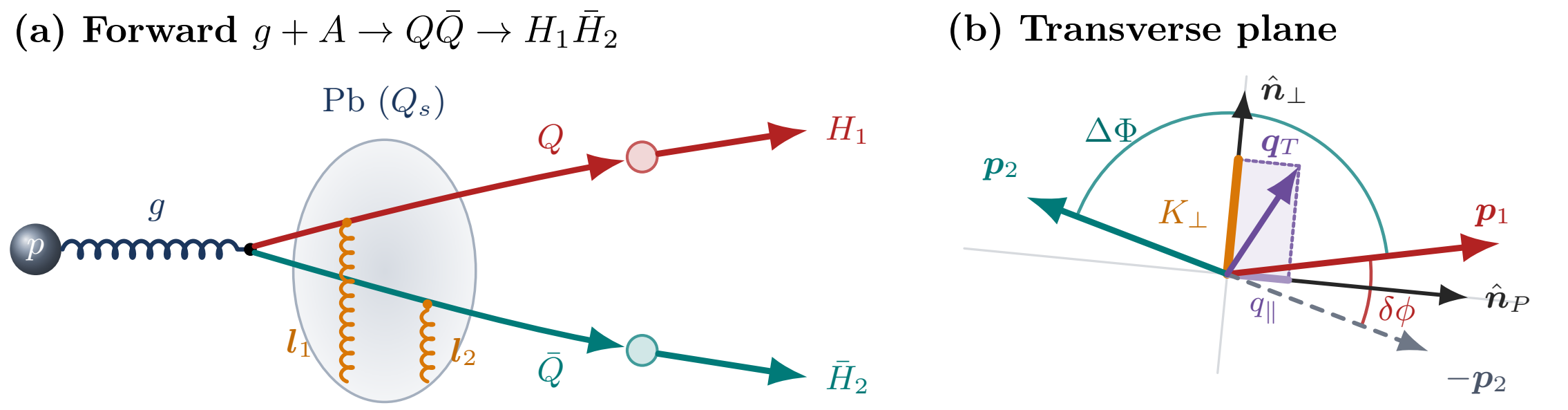
\begin{figure}[tbp]
  \centering
  \resizebox{0.98\textwidth}{!}{%
  \begin{tikzpicture}[
      >=Latex,
      line cap=round,
      line join=round,
      font=\small,
      gluon/.style={decorate, decoration={coil, aspect=0.60, segment length=4.6pt, amplitude=2.2pt}},
      kick/.style={decorate, decoration={coil, aspect=0.55, segment length=3.8pt, amplitude=1.6pt}},
    ]

    \node[anchor=west, font=\small\bfseries, text=black] at (0.0, 2.68) {(a) Forward $g + A \to Q\bar{Q} \to H_1\bar{H}_2$};

    \shade[inner color=jhepNavy!18, outer color=jhepNavy!5, draw=jhepNavy!40, line width=0.65pt]
      (3.60, 0.45) ellipse (0.85 and 1.22);
    \node[above, font=\footnotesize, text=jhepNavy] at (3.60, 1.72) {Pb ($Q_s$)};

    \shade[ball color=jhepNavy!62!white] (0.35, 0.65) circle (0.24);
    \node[font=\footnotesize\bfseries, text=white] at (0.35, 0.65) {$p$};
    \draw[gluon, jhepNavy, line width=1.15pt] (0.60, 0.65) -- (2.35, 0.65);
    \node[above, font=\footnotesize, text=jhepNavy] at (1.48, 0.78) {$g$};

    \fill[black] (2.35, 0.65) circle (0.06);
    \draw[->, line width=1.55pt, jhepCrimson]
      (2.39, 0.68) .. controls (3.55, 1.00) and (4.70, 1.26) .. (5.85, 1.48);
    \node[above, font=\footnotesize\bfseries, text=jhepCrimson] at (5.15, 1.40) {$Q$};
    \draw[->, line width=1.55pt, jhepTeal]
      (2.39, 0.62) .. controls (3.55, 0.28) and (4.70, -0.04) .. (5.85, -0.26);
    \node[below, font=\footnotesize\bfseries, text=jhepTeal] at (5.15, -0.18) {$\bar{Q}$};

    \coordinate (S1) at (3.25, -0.58);
    \coordinate (V1) at (3.25, 0.90);
    \coordinate (S2) at (4.00, -0.58);
    \coordinate (V2) at (4.00, 0.14);
    \draw[kick, jhepAmber, line width=1.0pt] (S1) -- (V1);
    \draw[kick, jhepAmber, line width=1.0pt] (S2) -- (V2);
    \fill[jhepAmber] (V1) circle (0.045);
    \fill[jhepAmber] (V2) circle (0.045);
    \node[left, font=\footnotesize, text=jhepAmber!90!black] at (3.08, -0.22) {$\bl_1$};
    \node[right, font=\footnotesize, text=jhepAmber!90!black] at (4.08, -0.30) {$\bl_2$};

    \draw[fill=jhepCrimson!18, draw=jhepCrimson!75, line width=0.75pt] (6.00, 1.51) circle (0.14);
    \draw[fill=jhepTeal!18, draw=jhepTeal!75, line width=0.75pt] (6.00, -0.29) circle (0.14);
    \draw[->, line width=1.65pt, jhepCrimson] (6.15, 1.54) -- (7.55, 1.76)
      node[right, font=\footnotesize] {$H_1$};
    \draw[->, line width=1.65pt, jhepTeal] (6.15, -0.32) -- (7.55, -0.54)
      node[right, font=\footnotesize] {$\bar{H}_2$};

    \begin{scope}[shift={(11.45, 0.42)}]
      \node[anchor=west, font=\small\bfseries, text=black] at (-2.75, 2.26) {(b) Transverse plane};

      \coordinate (O)      at (0.000, 0.000);
      \coordinate (P1)     at (2.550, 0.287);
      \coordinate (P2)     at (-1.880, 0.720);
      \coordinate (MP2)    at (1.880, -0.720);
      \coordinate (QT)     at (0.670, 1.007);
      \coordinate (NP)     at (2.250, -0.219);
      \coordinate (NPERP)  at (0.165, 1.720);
      \coordinate (QPAR)   at (0.569, -0.055);
      \coordinate (QPERP)  at (0.104, 1.067);

      \draw[jhepSlate!22, line width=0.6pt] (-2.45, 0.239) -- (2.75, -0.268);
      \draw[jhepSlate!22, line width=0.6pt] (-0.09, -0.92) -- (0.18, 1.84);
      \draw[jhepSlate!75, line width=0.65pt]
        (0.18, -0.017) -- (0.197, 0.162) -- (0.017, 0.179);

      \fill[jhepPurple!10] (O) -- (QPERP) -- (QT) -- (QPAR) -- cycle;
      \draw[densely dotted, jhepPurple!85, line width=0.9pt] (QT) -- (QPERP);
      \draw[densely dotted, jhepPurple!85, line width=0.9pt] (QT) -- (QPAR);

      \pic[draw=jhepTeal!75, line width=0.85pt, angle radius=1.50cm]
        {angle = P1--O--P2};
      \node[font=\footnotesize, text=jhepTeal!90!black] at (-1.08, 1.34) {$\Delta\Phi$};

      \pic[draw=jhepCrimson!85, line width=0.85pt, angle radius=1.34cm]
        {angle = MP2--O--P1};
      \node[font=\footnotesize, text=jhepCrimson] at (1.58, -0.35) {$\delta\phi$};

      \draw[->, line width=1.05pt, black!85] (O) -- (NP)
        node[right, font=\footnotesize] {$\hat{\bm{n}}_P$};
      \draw[->, line width=1.05pt, black!85] (O) -- (NPERP)
        node[right, font=\footnotesize] {$\hat{\bm{n}}_\perp$};

      \draw[line width=2.4pt, jhepAmber] (O) -- (QPERP);
      \node[left, font=\footnotesize\bfseries, text=jhepAmber!90!black] at (-0.02, 0.56) {$\Kperp$};

      \draw[line width=2.0pt, jhepPurple!60] (O) -- (QPAR);
      \node[below, font=\scriptsize, text=jhepPurple] at (0.35, -0.09) {$q_\parallel$};

      \draw[->, line width=1.65pt, jhepCrimson] (O) -- (P1)
        node[above, font=\footnotesize\bfseries, xshift=-2pt] {$\bp_1$};
      \draw[->, line width=1.65pt, jhepTeal] (O) -- (P2)
        node[above left, font=\footnotesize\bfseries, xshift=3pt] {$\bp_2$};
      \draw[->, dashed, line width=1.15pt, jhepSlate!80] (O) -- (MP2)
        node[below right, font=\footnotesize, text=jhepSlate] {$-\bp_2$};
      \draw[->, line width=1.55pt, jhepPurple] (O) -- (QT);
      \node[font=\footnotesize\bfseries, text=jhepPurple] at (0.50, 1.19) {$\bq_T$};
    \end{scope}
  \end{tikzpicture}}
  \caption{Production and transverse-plane measurement geometry for forward $g + A \to Q\bar{Q} \to H_1\bar{H}_2$ pair correlations. Panel~(a) illustrates the eikonal momentum transfers $\bl_1, \bl_2$ from the nuclear color field; panel~(b) defines the di-meson axis $\hat{\bm{n}}_P$, its orthogonal direction $\hat{\bm{n}}_\perp$, the pair imbalance $\bq_T$, and the orthogonal projection $\Kperp \equiv \aT = |\bq_T \cdot \hat{\bm{n}}_\perp|$.}
  \label{fig:geometry}
\end{figure}

Figure~\ref{fig:geometry} illustrates the space-time production picture and the transverse-plane measurement geometry. Without imposing a small-imbalance correlation expansion on the hard matrix element, the exact massive leading-order $g+A \to Q\bar{Q}$ hard factor retains the interference between gluon splitting before and after the target shockwave~\cite{Salazar:2026ssz}:
\begin{equation}
  \mathcal{H}_{gA}(\bk_1, \bk_2, \bl_1, \bl_2; z, m_Q)
  = \bm{t}^2\, \frac{\left[z^2 + (1-z)^2\right]\bm{h}^2 + m_Q^2 f^2}{\left[(\bk_1 - \bl_1)^2 + m_Q^2\right]^2},
  \label{eq:hard_gA}
\end{equation}
where $\bm{t} = z\bK' - \bl_1$, $\bm{d} = z\bK - \bl_1 = d\,\hat{\bm{u}}$, $f = [2(\bP\cdot\hat{\bm{u}}) + d]/(\bP^2 + m_Q^2)$, and $\bm{h} = f\bP - \hat{\bm{u}}$. Because $m_Q^2$ regulates the light-front energy denominators $(\bk_1 - \bl_1)^2 + m_Q^2$ and $\bP^2 + m_Q^2$, $\mathcal{H}_{gA}$ remains finite and positive across the entire azimuthal domain $\Delta\Phi \in [0, \pi]$, including the collinear near-side limit $\Delta\Phi \to 0$. At the same time, the large mass ratio $m_b^2 / m_c^2 \approx 12.4$ suppresses near-side $g \to b\bar{b}$ splittings relative to $g \to c\bar{c}$, which directly shapes the away-to-near-side balance in LHCb correlation measurements~\cite{LHCb:2012aiv,LHCb:2017bmm}.

In the back-to-back region where the net radiative recoil $\bq_{\mathrm{rad}} = \bK - \bK'$ is small compared with $M_{\mathrm{inv}}$, soft and collinear gluon emissions generate large Sudakov logarithms that must be resummed in conjugate impact-parameter space $\bb$. Away from the back-to-back region, where the target momentum transfer $|\bK'|$ becomes comparable to $M_{\mathrm{inv}}$, the logarithmic hierarchy breaks down and the cross section should smoothly recover the leading-order kinematic-twist behavior without artificial perturbative Sudakov damping. Following ref.~\cite{Salazar:2026ssz} (see also refs.~\cite{Mueller:2013wwa,Watanabe:2015yca,Shi:2021hwx,Wang:2022fwj}), we implement a smooth threshold-matching prescription:
\begin{equation}
  W_{\mathrm{match}}(q) = g_{\mathrm{match}}\, W_{\mathrm{soft}}(q) + (1 - g_{\mathrm{match}})\, W_{\mathrm{hard}}(q),
  \qquad g_{\mathrm{match}} = \frac{M_{\mathrm{inv}}^2}{M_{\mathrm{inv}}^2 + c_{\mathrm{match}}^2 \bK'^2},
  \label{eq:W_match}
\end{equation}
with nominal matching parameter $c_{\mathrm{match}} = 1.0$ and variation interval $c_{\mathrm{match}} \in \{0.5, 2.0\}$. The resummed soft-collinear kernel and the large-angle matching kernel are given by
\begin{align}
  W_{\mathrm{soft}}(q) &= \int_0^\infty \frac{b\,\dd b}{2\pi}\, J_0(q b)\, x_p g_p(x_p, \mu_b)\,
    D_{H/Q}(\xi_1, \mu_b)\, D_{H/Q}(\xi_2, \mu_b) \nonumber\\
  &\quad \times e^{-S_P(b;\mu, m_Q) - S_{\mathrm{NP}}(b;\mu,\CNP)},
  \label{eq:W_soft} \\
  W_{\mathrm{hard}}(q) &= x_p g_p(x_p, \mu)\, D_{H/Q}(\xi_1, \mu)\, D_{H/Q}(\xi_2, \mu)
    \int_0^\infty \frac{b\,\dd b}{2\pi}\, J_0(q b)\, e^{-S_{\mathrm{NP}}(b;\mu,\CNP)},
  \label{eq:W_hard}
\end{align}
where $\mu_b = \min(\mu, c_0/b_*)$ is the CSS impact-parameter scale with $c_0 = 2e^{-\gamma_E}$, $b_* = b/\sqrt{1 + b^2/b_{\max}^2}$, and $b_{\max} = 1.5\gev^{-1}$~\cite{Collins:1984kg}. In $W_{\mathrm{soft}}(q)$, evaluating the projectile gluon PDF and heavy-meson fragmentation functions at $\mu_b$ resums the DGLAP collinear logarithms associated with initial- and final-state radiation up to the hard scale $\mu$. In the large-$N_c$ dipole framework, finite-$N_c$ quadrupole corrections and color-correlated soft anomalous dimensions enter at order $1/N_c^2$; for self-normalized away-side ratios at $p_T \ge 4\gev$, the leading nuclear modification is controlled by the target saturation scale $\Qs(x_A)$.

\subsection{SCET/bHQET dead-cone Sudakov factor}
\label{sec:deadcone}

The perturbative Sudakov exponent $S_P(b; \mu, m_Q)$ in eq.~\eqref{eq:W_soft} governs how soft-collinear radiation depends on the heavy-quark mass $m_Q$. Within SCET/bHQET factorization for massive $g+A \to Q\bar{Q}+X$ production~\cite{Fleming:2007qr,Fleming:2007xt,Korchemsky:1987wg,Salazar:2026ssz}, the renormalization-group evolution between $\mu_b$ and $\mu$ separates at the heavy-quark mass threshold $\bar{\mu} = m_Q$, yielding
\begin{align}
  S_P(b; \mu, m_Q) = \int_{\mu_b^2}^{\mu^2} \frac{\dd\bar{\mu}^2}{\bar{\mu}^2} \frac{\alpha_s(\bar{\mu})}{2\pi} \bigg\{
  & (2C_A + 2C_F)\ln\frac{\mu^2}{\bar{\mu}^2} - 2\beta_0(n_f) - 3 C_F \nonumber\\
  & + \Theta(m_Q^2 - \bar{\mu}^2) \left[ C_F - 2 C_F \ln\frac{m_Q^2}{\bar{\mu}^2} \right] \bigg\},
  \label{eq:SP_deadcone}
\end{align}
where $\beta_0(n_f) = (11C_A - 2n_f)/12$ with active flavor thresholds at $m_c$ and $m_b$. Above the mass threshold ($\bar{\mu} > m_Q$), the heavy quarks radiate as effectively massless color charges with total cusp anomalous dimension $2C_A + 2C_F$. Below the mass threshold ($\bar{\mu} < m_Q$), the step function switches the final-state evolution from SCET to bHQET: the final-state cusp double logarithm freezes at $2C_F \ln(\mu^2 / m_Q^2)$ while the single-logarithmic noncusp anomalous dimension shifts from $-3C_F$ to $-2C_F$. This piecewise structure is the impact-parameter representation of the QCD dead cone.

To see how the dead-cone threshold affects $S_P(b; \mu, m_Q)$ across charm and bottom flavors, we differentiate eq.~\eqref{eq:SP_deadcone} with respect to $m_Q^2$ in the active regime $\mu_b < m_Q < \mu$:
\begin{equation}
  \frac{\partial S_P(b; \mu, m_Q)}{\partial m_Q^2}
  = \frac{C_F}{m_Q^2}\left[ \frac{\alpha_s(m_Q)}{2\pi}
    - 2\int_{\mu_b^2}^{m_Q^2} \frac{\dd\bar{\mu}^2}{\bar{\mu}^2}\frac{\alpha_s(\bar{\mu})}{2\pi} \right],
  \qquad (\mu_b < m_Q < \mu),
  \label{eq:deadcone_deriv}
\end{equation}
while $\partial S_P/\partial m_Q^2 = 0$ for $m_Q \le \mu_b$. Equation~\eqref{eq:deadcone_deriv} reflects the competition between the positive single-logarithmic noncusp shift $+C_F$ (first term, coming from the upper boundary of $\Theta(m_Q^2 - \bar{\mu}^2)$) and the negative double-logarithmic cusp screening $-2C_F \ln(m_Q^2/\bar{\mu}^2)$ (second term). In the fixed-coupling approximation, integrating the mass shift $\Delta S_P \equiv S_P(m_Q) - S_P(0)$ over $\Delta t \equiv \ln(m_Q^2/\mu_b^2) > 0$ gives
\begin{equation}
  \Delta S_P \simeq \frac{\alpha_s C_F}{2\pi}\, \Delta t\,(1 - \Delta t),
  \label{eq:deadcone_threshold}
\end{equation}
which turns negative, corresponding to a net suppression of perturbative Sudakov radiation relative to the massless limit, once the heavy-quark mass exceeds the crossover scale
\begin{equation}
  \Delta t > 1 \quad \Longleftrightarrow \quad m_Q > e^{1/2}\,\mu_b \approx 1.649\,\mu_b.
  \label{eq:deadcone_onset}
\end{equation}
Because the CSS $b_*$ prescription bounds $\mu_b$ from below by $\mu_b^{\min} = c_0 / b_{\max} = 0.749\gev$, this fixed-coupling crossover occurs at $e^{1/2}\mu_b^{\min} \simeq 1.23\gev$. Charm ($m_c = 1.28\gev$) sits right at this boundary and therefore experiences almost no net dead-cone reduction of $S_P$, whereas bottom ($m_b = 4.50\gev$) lies well above $e^{1/2}\mu_b$ for all $b \gtrsim 0.43\gev^{-1}$, resulting in a substantial reduction of perturbative Sudakov broadening. In all numerical calculations below, eq.~\eqref{eq:SP_deadcone} is integrated with the exact running coupling.

At large impact parameters $b \sim 1\gev^{-1}$, the perturbative exponent is supplemented by the nonperturbative CSS Sudakov factor~\cite{Sun:2014dqm,Salazar:2026ssz}
\begin{equation}
  S_{\mathrm{NP}}(b; \mu, \CNP) = \CNP \left(\frac{C_A}{C_F} + 2\right)
  \left[ \frac{g_1}{2} b^2 + \frac{g_2}{8} \ln\!\left(\frac{\mu^2}{Q_0^2}\right) \ln\!\left(1 + \frac{b^2}{b_{\max}^2}\right) \right],
  \label{eq:SNP}
\end{equation}
with $g_1 = 0.212\gev^2$, $g_2 = 0.84$, $Q_0^2 = 2.4\gev^2$, and baseline normalization $\CNP = 2.3$. Because $\CNP$ is not yet tightly constrained for massive final states in forward collisions, we propagate a five-point scan $\CNP \in \{1.5, 1.9, 2.3, 2.9, 3.5\}$ through all observables in section~\ref{sec:results}.

\subsection{Nuclear geometry, opacity scaling, and non-saturation baselines}
\label{sec:baselines}

To determine whether an observed nuclear broadening requires nonlinear gluon saturation or can be explained by conventional cold-nuclear-matter mechanisms, we evaluate three distinct target descriptions within the identical hard-scattering, fragmentation, and Sudakov framework.

For our baseline CGC saturation predictions, we use the Balitsky running-coupling BK-evolved AAMQS MV-$e$ proton dipole $S_p(Y, r)$ and impact-parameter-dependent lead dipole $S_{\mathrm{Pb}}(Y, r; b_\perp)$ constrained by HERA inclusive structure-function data~\cite{Albacete:2010sy,Lappi:2013zma,Casuga:2023dcf,Dai:2026dlq,Dai:2026ddp}. At evolution rapidity $Y = 4.4$ ($x_A \simeq 1.2\times 10^{-4}$), these tables correspond to fundamental saturation scales $\Qs^2 = 0.534\gev^2$ for the proton and $1.734\gev^2$ for central ($b_\perp = 0$) $\pPb$, where $\Qs^2(Y) \equiv 2 / r_s^2(Y)$ is defined by $S_A(Y, r_s) = e^{-1/2}$. For minimum-bias $\pPb$ collisions, we construct the binary-collision-weighted impact-parameter average $S_{\mathrm{Pb}}^{\mathrm{MB}}(Y, r) = \langle S_{\mathrm{Pb}}(Y, r; b_\perp)\rangle_{b_\perp}$ using a standard Woods--Saxon nuclear thickness profile $T_{\mathrm{Pb}}(b_\perp)$ ($R_{\mathrm{Pb}} = 6.62\,\mathrm{fm}$, $a_{\mathrm{Pb}} = 0.546\,\mathrm{fm}$). This minimum-bias dipole yields $\Qs^2 = 1.256\gev^2$ at $Y = 4.4$ and enters eq.~\eqref{eq:master_xsec} as an effective mean-field dipole amplitude. The pre-tabulated AAMQS grids extend up to $Y = 5.6$ ($x_A \ge 3.7\times 10^{-5}$); this range covers the $[4,6]\gev$ and $[6,8]\gev$ windows completely (where $\langle x_A\rangle \simeq (7.0\text{--}11.5)\times 10^{-5}$), while freezing the ultra-small-$x_A$ integrand tail at $Y = 5.6$ in the lowest $[2,4]\gev$ bin.

In addition to fixed $\pp$ and $\pPb$ comparisons, the LHC Run~3--4 program includes proton-oxygen ($\pO$) collisions, which bridge the gap between proton and heavy-nuclear targets. To map the continuous dependence of our observables on the nuclear saturation scale and to quantify the theoretical uncertainty associated with the initial nuclear saturation scale $Q_{s0,\mathrm{Pb}}^2$, we solve the one-dimensional Balitsky rcBK equation across eight initial opacity ratios $\nu_A \equiv Q_{s0,A}^2/Q_{s0,p}^2 \in [1.0, 4.33]$. Matching the evolved $1\mathrm{D}$ rcBK saturation scale to the central-Pb AAMQS table fixes the nominal lead opacity ratio to $\nu_{\mathrm{Pb},0} = 3.608$, so $\pm 20\%$ variations of $Q_{s0,\mathrm{Pb}}^2$ correspond to $\nu_A = 2.886$ and $4.330$. For $0\text{--}10\%$ central $\pO$ collisions, we determine the impact-parameter cutoff from the Glauber inelastic probability $P_{pA}(b) = 1 - \exp[-\sigma_{NN}^{\mathrm{inel}} T_A(b)]$ with Woods--Saxon parameters $R_{\mathrm{O}} = 2.608\,\mathrm{fm}$, $a_{\mathrm{O}} = 0.513\,\mathrm{fm}$, and $\sigma_{NN}^{\mathrm{inel}} = 70\,\mathrm{mb}$. Taking the binary-collision-weighted average inside the $0\text{--}10\%$ centrality class gives $\nu_{\mathrm{O},0\text{--}10\%} = \nu_{\mathrm{Pb},0}\langle T_{\mathrm{O}}\rangle_{\mathrm{coll},0\text{--}10\%}/T_{\mathrm{Pb}}(0) = 1.3054$, corresponding to $\Qs^2(Y=4.4) = 0.666\gev^2$.

We contrast these nonlinear CGC predictions against two non-saturation baselines. The first baseline isolates collinear nuclear shadowing without nonlinear transverse-momentum broadening: we retain the proton dipole distribution $F_p(x_A, \bl)$ in eq.~\eqref{eq:master_xsec} and reweight the soft and hard Sudakov kernels by the central \texttt{nNNPDF3.0} nuclear gluon modification ratio~\cite{AbdulKhalek:2022fyi},
\begin{equation}
  R_{g,\mathrm{soft}}^{\mathrm{Pb}} = R_g^{\mathrm{Pb}}\!\left(x_A,\max(\mu_b,2\gev)\right),
  \qquad
  R_{g,\mathrm{hard}}^{\mathrm{Pb}} = R_g^{\mathrm{Pb}}\!\left(x_A,\max(\mu,2\gev)\right).
  \label{eq:npdf_base}
\end{equation}
Because $R_g^{\mathrm{Pb}}(x_A, \mu)$ increases with the scale $\mu$ due to DGLAP evolution, this collinear baseline tests whether scale-dependent shadowing of the longitudinal momentum fraction $x_A$ across the pair phase space can mimic transverse-momentum broadening in self-normalized ratios. The second baseline represents linear Gaussian multiple scattering without nonlinear rcBK evolution: we again start from the proton dipole table and multiply the coordinate-space dipole by a Gaussian broadening factor,
\begin{equation}
  \exp\!\left[-\frac{1}{4}\Delta_{\mathrm{Pb}}^2(x_A)\, b^2\right],
  \qquad \Delta_{\mathrm{Pb}}^2(x_A) = Q_{s,\mathrm{Pb}}^2(x_A) - Q_{s,p}^2(x_A).
  \label{eq:gauss_base}
\end{equation}
Although eq.~\eqref{eq:gauss_base} has the same small-$b$ quadratic saturation-scale shift $\Delta_{\mathrm{Pb}}^2(x_A)$ as the central-Pb CGC dipole, a pure Gaussian decays exponentially at large transverse momentum $l$, whereas the rcBK-evolved nuclear dipole develops a power-law semi-hard tail $F_A(x_A, \bl) \propto \Qs^2 / l^4$ for $l \gtrsim \Qs$. Comparing the full CGC prediction against eq.~\eqref{eq:gauss_base} therefore tests whether the away-side broadening is driven merely by a soft Gaussian shift at $l \lesssim 1\gev$ or by coherent scatterings in the semi-hard $l \sim 2\text{--}6\gev$ tail of the rcBK dipole.

\section{Heavy-flavor-tagged TEEC and DPS cancellation}
\label{sec:teec}

With the massive hybrid-CGC cross section and target baselines in place, we now construct the angular energy-correlator observables designed to enhance the parent heavy-quark kinematics and suppress uncorrelated multiparton backgrounds.

\subsection{\texorpdfstring{Definitions of $\SigmaHF^{(n)}(\tau)$ and heavy-flavor ATEEC}{Definitions of Sigma HF moments and heavy-flavor ATEEC}}

Let $\Delta\Phi = \phi_1 - \phi_2 \in [-\pi, \pi]$ denote the azimuthal angle between the reconstructed heavy mesons $H_1$ and $\bar{H}_2$, and let $\delta\phi \equiv \pi - |\Delta\Phi| \in [0, \pi]$ measure the acoplanarity angle relative to the back-to-back configuration. Following standard hadron-collider TEEC conventions~\cite{Gao:2019ojf,Kang:2024kpz,Kang:2025kklp,Kang:2026klp}, we map the azimuthal separation onto the dimensionless variable
\begin{equation}
  \tau \equiv \frac{1 + \cos\Delta\Phi}{2} = \sin^2\!\left(\frac{\delta\phi}{2}\right) \in [0, 1],
  \qquad \frac{\dd\tau}{\dd\delta\phi} = \sqrt{\tau(1-\tau)},
  \label{eq:tau_def}
\end{equation}
so that the back-to-back region $\Delta\Phi \to \pi$ corresponds to $\tau \to 0$ ($\tau \approx \delta\phi^2 / 4$ for $\delta\phi \ll 1$), the orthogonal configuration $\Delta\Phi = \pi/2$ corresponds to $\tau = 1/2$, and the collinear near-side limit $\Delta\Phi \to 0$ corresponds to $\tau \to 1$. Within a fiducial acceptance $\Omega$ defined by $p_{T1}, p_{T2} \in [p_{T,\min}, p_{T,\max}]$ and $y_1, y_2 \in [y_{\min}, y_{\max}]$, we define the order-$n$ heavy-flavor-tagged transverse energy--energy correlator (HF-TEEC) by weighting each pair with the $n$-th power of the meson transverse energies $E_{Ti} = \sqrt{p_{Ti}^2 + M_H^2}$:
\begin{equation}
  \SigmaHF^{(n)}(\tau) \equiv \frac{1}{\mathcal{N}_{\mathrm{HF}}^{(n)}}
  \int_\Omega \dd\mathcal{P}_1\,\dd\mathcal{P}_2\, (E_{T1} E_{T2})^n
  \frac{\dd\sigma^{pA\to H\bar{H}X}}{\dd\mathcal{P}_1\,\dd\mathcal{P}_2}\,
  \delta\!\left(\tau - \frac{1 + \cos\Delta\Phi}{2}\right),
  \label{eq:HF_TEEC_def}
\end{equation}
where $\dd\mathcal{P}_i = \dd^2\bp_i\,\dd y_i$ and $\mathcal{N}_{\mathrm{HF}}^{(n)}$ is the normalization integral. Setting $n = 0$ recovers the unweighted azimuthal correlation~\cite{Salazar:2026ssz}, $n = 1$ gives the standard linear transverse-energy weight, and $n = 2$ defines a quadratic energy weight. It is instructive to contrast eq.~\eqref{eq:HF_TEEC_def} with inclusive all-hadron TEEC definitions~\cite{Kang:2025kklp,Kang:2026klp}: in the latter, summing over all unidentified hadrons and integrating over $0 < p_{Ti} < \infty$ invokes the collinear momentum sum rule $\sum_h \int_0^1 \dd\xi\,\xi D_{h/Q}(\xi,\mu) = 1$ to cancel collinear fragmentation functions at leading order. In forward LHCb measurements, however, one tags exclusively reconstructed $\DD$ or $\BB$ pairs inside finite $p_T$ intervals. Rather than eliminating fragmentation via an all-hadron sum, eq.~\eqref{eq:HF_TEEC_def} uses the $(E_{T1}E_{T2})^n$ weight to shift the sampled heavy-quark fragmentation distribution toward harder momentum fractions while retaining access to the Sudakov-resummed back-to-back region $\tau \in [0, 0.5]$. Self-normalizing $\SigmaHF^{(n)}(\tau)$ on the away-side hemisphere $\int_0^{1/2}\dd\tau\,\SigmaHF^{(n)}(\tau) = 1$ (or over $\tau \in [0,1]$ when comparing full-azimuthal shapes) cancels overall integrated luminosity, reconstruction efficiency, and branching-fraction uncertainties.

To separate the asymmetric away-side recoil peak from azimuthally symmetric backgrounds and near-side collinear splittings, we further define the mirror-antisymmetric Heavy-Flavor ATEEC (HF-ATEEC) on the away-side domain $\tau \in [0, 1/2]$:
\begin{equation}
  A_{\mathrm{HF}}^{(n)}(\tau) \equiv \SigmaHF^{(n)}(\tau) - \SigmaHF^{(n)}(1 - \tau).
  \label{eq:ATEEC_def}
\end{equation}
Because $\tau \mapsto 1 - \tau$ corresponds to reflecting $\Delta\Phi \mapsto \pi - \Delta\Phi$ across $\Delta\Phi = \pi/2$, eq.~\eqref{eq:ATEEC_def} can be formed directly from any symmetrically binned experimental $\Delta\Phi \in [0, \pi]$ histogram via $\sqrt{\tau(1-\tau)}\,A_{\mathrm{HF}}^{(n)}(\tau) = \frac{1}{\mathcal{N}}\frac{\dd\Sigma^{(n)}}{\dd\Delta\Phi}(\pi - \delta\phi) - \frac{1}{\mathcal{N}}\frac{\dd\Sigma^{(n)}}{\dd\Delta\Phi}(\delta\phi)$ without requiring an unbinned re-analysis.

\subsection{Mellin-moment fragmentation hardening}

We next quantify how the transverse-energy weight $(E_{T1}E_{T2})^n$ in eq.~\eqref{eq:HF_TEEC_def} modifies the sampled fragmentation fractions $\xi_1, \xi_2$. At fixed parent heavy-quark transverse momentum $k_T$, the meson transverse energy $E_T(\xi, k_T) \simeq \sqrt{\xi^2 k_T^2 + M_H^2}$ is a strictly increasing function of $\xi$, with logarithmic derivative $\partial\ln E_T / \partial\xi = \xi k_T^2 / E_T^2 > 0$. Differentiating the energy-weighted expectation value of the fragmentation fraction with respect to the weight index $n$ yields the exact cumulant relations
\begin{equation}
  \left.\frac{\dd\langle\xi\rangle_n}{\dd n}\right|_{k_T} = \operatorname{Cov}_n(\xi, \ln E_T) > 0,
  \qquad
  \frac{\dd\langle\bar{\xi}\rangle_n}{\dd n} = \operatorname{Cov}_n\!\left(\bar{\xi},\, \ln(E_{T1}E_{T2})\right),
  \label{eq:thm1}
\end{equation}
where the first relation holds unconditionally at fixed parton $k_T$ for any non-degenerate fragmentation function $D_{H/Q}(\xi, \mu)$, and the second governs the pair-averaged fraction $\bar{\xi} = \frac{1}{2}(\xi_1+\xi_2)$ averaged over a finite hadronic acceptance $\Omega$. In the high-$p_T$ regime at fixed parton kinematics, $(E_{T1}E_{T2})^n \propto (\xi_1\xi_2)^n$ shifts the effective Mellin moment $\mathcal{M}_n(\mu) = \int_0^1 \dd\xi\,\xi^n D_{H/Q}(\xi, \mu)$ upward by $+n$, suppressing the soft-$\xi$ tail of the fragmentation function. Inside a fixed hadron-$p_T$ bin $[p_{T,\min}, p_{T,\max}]$, the pair covariance $\operatorname{Cov}_n(\bar{\xi}, \ln(E_{T1}E_{T2}))$ remains positive whenever the underlying partonic cross section steepens with increasing $p_T$ above the mass and saturation scales, because higher meson transverse energies $E_{Ti}$ within the bin require larger fragmentation fractions $\xi_i$ to avoid paying the steep partonic spectrum penalty at large $k_{Ti} = \lambda_i p_{Ti}$. This hardening is most pronounced for charm pairs in wide transverse-momentum windows (such as $2 < p_T < 8\gev$), where $D_{D/c}(\xi)$ is broad and $E_T$ varies by more than a factor of three across the bin; it is smaller for bottom pairs, where $D_{B/b}(\xi)$ is already narrowly peaked near $\xi \simeq 0.80\text{--}0.85$, and inside narrow $2\gev$-wide bins.

\subsection{DPS pedestal suppression and exact ATEEC cancellation}

A major experimental challenge in forward $\pPb$ heavy-flavor correlations is the large pedestal generated by Double Parton Scattering (DPS). When two independent hard gluon-fusion interactions occur within the same proton-nucleus collision, they produce an uncorrelated heavy-meson pair whose folded azimuthal distribution $|\Delta\Phi| \in [0, \pi]$ is flat~\cite{Strikman:2001gz,dEnterria:2012jam,Helenius:2019uge,Blok:2019fgg,LHCb:2020jse}:
\begin{equation}
  \frac{\dd\sigma_{\mathrm{DPS}}^{pA}}{\dd p_{T1}\,\dd y_1\,\dd p_{T2}\,\dd y_2\,\dd|\Delta\Phi|}
  = \frac{F_{\mathrm{DPS}}^{pA}}{\pi\,A\,\sigma_{\mathrm{eff}}^{pp}}
  \frac{\dd\sigma_{\mathrm{trig}}^{pA}}{\dd p_{T1}\,\dd y_1}
  \frac{\dd\sigma_{\mathrm{trig}}^{pA}}{\dd p_{T2}\,\dd y_2}.
  \label{eq:DPS_def}
\end{equation}
In proton-nucleus collisions, the two partons from the projectile proton can scatter either off a single target nucleon or off two distinct nucleons at the same impact parameter, enhancing the per-nucleon DPS cross section by $F_{\mathrm{DPS}}^{p\mathrm{Pb}} = 1 + \sigma_{\mathrm{eff}}^{pp}\frac{A-1}{A}\langle T_A\rangle \simeq 3\text{--}4$ for $\sigma_{\mathrm{eff}}^{pp} = 15\,\mathrm{mb}$. This nuclear enhancement accounts for the large mid-angle pedestal observed by LHCb in $\pPb$ open-charm production~\cite{LHCb:2020jse}.

Transverse-energy weighting and mirror antisymmetrization suppress this DPS contribution in two complementary ways. First, let $f_{\mathrm{DPS}}^{(n)}$ denote the ratio of the energy-weighted DPS yield to the single parton scattering (SPS) yield inside the acceptance $\Omega$. Because the DPS cross section~\eqref{eq:DPS_def} factorizes into a product of single-inclusive trigger spectra, the relative DPS contamination at weight order $n$ satisfies
\begin{equation}
  \frac{f_{\mathrm{DPS}}^{(n)}}{f_{\mathrm{DPS}}^{(0)}}
  = \frac{\langle E_{T1}^n \rangle_{\mathrm{trig}}\,\langle E_{T2}^n \rangle_{\mathrm{trig}}}{\langle (E_{T1}E_{T2})^n \rangle_{\mathrm{SPS}}}.
  \label{eq:thm2a}
\end{equation}
In SPS events, recoil conservation correlates the two meson transverse momenta ($p_{T1} \approx p_{T2}$), so both mesons populate the high-$E_T$ tail together and generate a positive pair covariance $\langle E_{T1}^n E_{T2}^n\rangle_{\mathrm{SPS}} > \langle E_{T1}^n\rangle_{\mathrm{trig}}\langle E_{T2}^n\rangle_{\mathrm{trig}}$. In DPS events, by contrast, the two mesons are drawn independently from the steeply falling single-inclusive spectrum, so configurations with two simultaneous high-$p_T$ mesons are strongly suppressed. Consequently, raising $n$ in eq.~\eqref{eq:thm2a} systematically reduces the relative DPS background across wide $p_T$ intervals.

Second, while energy weighting reduces the DPS fraction in $\SigmaHF^{(n)}(\tau)$, the mirror-antisymmetric observable $A_{\mathrm{HF}}^{(n)}(\tau)$ removes any azimuthally flat DPS pedestal—together with all even azimuthal harmonics $\cos(2m\Delta\Phi)$ from elliptic flow or soft background fluctuations—at the algebraic level. Because the coordinate Jacobian $\dd\tau / \dd\delta\phi = \sqrt{\tau(1-\tau)}$ in eq.~\eqref{eq:tau_def} is invariant under the reflection $\tau \mapsto 1 - \tau$, any component of $\dd\sigma/\dd\Delta\Phi$ that is symmetric under $\Delta\Phi \mapsto \pi - \Delta\Phi$ cancels identically in eq.~\eqref{eq:ATEEC_def}:
\begin{equation}
  A_{\mathrm{DPS}+\text{even}}^{(n)}(\tau) \equiv 0 \qquad \forall\,\tau \in \left(0, \tfrac{1}{2}\right].
  \label{eq:thm2b}
\end{equation}
Furthermore, because $A_{\mathrm{HF}}^{(n)}(\tau)$ measures the difference between the away-side ($\tau < 1/2$) and near-side ($1-\tau > 1/2$) hemispheres, its sign provides a direct diagnostic of the production mechanism: for bottom pairs, the strong mass suppression of near-side $g \to b\bar{b}$ splitting keeps $A_{\mathrm{HF}}^{(n)}(\tau) > 0$ across the entire away-side domain, whereas for charm pairs the narrower near-side collinear peak drives $A_{\mathrm{HF}}^{(n)}(\tau)$ negative as $\tau \to 0$.

\section{\texorpdfstring{Single-axis projections ($\Kperp \equiv \aT$) and Fourier-cosine transforms}{Single-axis projections (Kperp equals aT) and Fourier-cosine transforms}}
\label{sec:proj}

The angular observables $\SigmaHF^{(n)}(\tau)$ and $A_{\mathrm{HF}}^{(n)}(\tau)$ formulated in section~\ref{sec:teec} operate purely in azimuthal angle space. Inside a finite transverse-momentum bin $[p_{T,\min}, p_{T,\max}]$, a fixed acoplanarity angle $\delta\phi$ mixes pairs of different absolute transverse momenta and therefore does not measure the target momentum transfer $|\bK'|$ directly in GeV units. Conversely, the total two-dimensional pair imbalance $|\bq_T| = |\bp_1 + \bp_2|$ is heavily distorted by longitudinal fragmentation recoil. In this section, we resolve this tension by introducing the single-axis orthogonal projection $\Kperp \equiv \aT$ and its one-dimensional Fourier-cosine transform.

\subsection{Kinematic decomposition and suppression of longitudinal fragmentation recoil}

Following the Thrust-axis projection originally introduced for Drell--Yan pairs in ref.~\cite{Banfi:2010xy}, we define the di-meson thrust axis $\bP_h = \frac{1}{2}(\bp_1 - \bp_2)$ with orthonormal unit vectors $\hat{\bm{n}}_P = \bP_h / |\bP_h|$ and $\hat{\bm{n}}_\perp = \hat{\bm{z}} \times \hat{\bm{n}}_P$ in the transverse plane (shown schematically in figure~\ref{fig:geometry}(b)). Decomposing the total pair transverse-momentum imbalance $\bq_T = \bp_1 + \bp_2$ along and orthogonal to $\hat{\bm{n}}_P$ yields
\begin{align}
  \Kperp \equiv \aT \equiv |\bq_T \cdot \hat{\bm{n}}_\perp|
  &= \frac{2 p_{T1} p_{T2} |\sin\Delta\Phi|}{\sqrt{p_{T1}^2 + p_{T2}^2 - 2p_{T1}p_{T2}\cos\Delta\Phi}}, \nonumber\\
  q_\parallel \equiv \bq_T \cdot \hat{\bm{n}}_P
  &= \frac{p_{T1}^2 - p_{T2}^2}{\sqrt{p_{T1}^2 + p_{T2}^2 - 2p_{T1}p_{T2}\cos\Delta\Phi}},
  \label{eq:Kperp_def}
\end{align}
which satisfy the exact Pythagorean decomposition $q_\parallel^2 + \Kperp^2 = |\bq_T|^2$. To benchmark $\Kperp \equiv \aT$ against traditional correlation variables, we also evaluate the trigger-orthogonal out-of-plane momentum $p_{\mathrm{out}} \equiv p_{T2}|\sin\Delta\Phi|$.

To see why $\Kperp \equiv \aT$ isolates the transverse recoil from collinear fragmentation smearing, consider a nearly back-to-back parent heavy-quark pair ($k_{T1} = k_{T2} = k_T$, $\delta\phi \ll 1$) whose daughters fragment with unequal momentum fractions $\xi_1 = \bar{\xi}(1 + \epsilon)$ and $\xi_2 = \bar{\xi}(1 - \epsilon)$, parameterized by the relative fragmentation asymmetry $\epsilon \equiv (\xi_1 - \xi_2)/(\xi_1 + \xi_2) \in (-1, 1)$. Expanding the denominator $|\bp_1 - \bp_2| = 2\bar{\xi}k_T[1 - \frac{1}{4}(1-\epsilon^2)\delta\phi^2 + \mathcal{O}(\delta\phi^4)]$ together with $2p_{T1}p_{T2}\sin\delta\phi = 2\bar{\xi}^2(1-\epsilon^2)k_T^2\delta\phi + \mathcal{O}(\delta\phi^3)$ and $p_{T1}^2 - p_{T2}^2 = 4\bar{\xi}^2\epsilon k_T^2$ gives the small-angle expansions
\begin{align}
  q_\parallel &= 2\bar{\xi} k_T\,\epsilon + \mathcal{O}(\delta\phi^2), \label{eq:qpar_noise}\\
  p_{\mathrm{out}} &= \bar{\xi}(1-\epsilon) k_T\,\delta\phi + \mathcal{O}(\delta\phi^3), \label{eq:pout_noise}\\
  \Kperp \equiv \aT &= \bar{\xi}(1-\epsilon^2) k_T\,\delta\phi + \mathcal{O}(\delta\phi^3). \label{eq:Kperp_noise}
\end{align}
Equations~\eqref{eq:qpar_noise}--\eqref{eq:Kperp_noise} reveal a sharp structural difference among the three momentum-space variables. Because $q_\parallel$ is proportional to $\epsilon$ even at $\delta\phi = 0$, any event-by-event fluctuation in $\xi_1 - \xi_2$ generates a large longitudinal imbalance $q_\parallel \sim 1\text{--}2\gev$. In the two-dimensional vector sum $|\bq_T| = \sqrt{q_\parallel^2 + \Kperp^2}$ (and its radial phase-space Jacobian $|\bq_T|\,\dd|\bq_T|$), this $\mathcal{O}(\epsilon)$ longitudinal noise completely overwhelms the small-$\Kperp$ region, forcing $\dd\sigma/\dd|\bq_T|$ to vanish at $|\bq_T| = 0$ and shifting its peak to $|\bq_T| \simeq 1.5\gev$. The trigger-projected variable $p_{\mathrm{out}}$ avoids the $\delta\phi = 0$ longitudinal offset, but still suffers from linear $\mathcal{O}(\epsilon)$ fragmentation sensitivity and develops a sharp square-root Jacobian singularity at the bin boundary $p_{\mathrm{out}} = p_{T,\min}$ where $|\sin\Delta\Phi| = 1$. By contrast, the symmetric factor $2p_{T1}p_{T2}/|\bp_1 - \bp_2|$ in $\Kperp \equiv \aT$ is an even function of $\epsilon$: its first derivative $\partial\Kperp/\partial\epsilon|_{\epsilon=0} = 0$ vanishes identically, so fragmentation fluctuations enter only at quadratic order $\mathcal{O}(\epsilon^2)$. Because heavy-quark fragmentation functions are already narrow and $(E_{T1}E_{T2})^n$ weighting further reduces the variance of $\epsilon$, $\Kperp$ provides a smooth, cusp-free spectrum peaked at $\Kperp = 0$ that directly tracks the transverse recoil $\bar{\xi} k_T \delta\phi$.

Projecting onto the orthogonal axis also simplifies the conjugate impact-parameter structure. Integrating the two-dimensional Fourier transform of an isotropic $b$-space kernel $\widetilde{\mathcal{W}}(b)$ over the unmeasured longitudinal component $q_\parallel$ projects the two-dimensional $J_0(|\bq_T|b)$ Hankel transform onto a one-dimensional Fourier-cosine transform~\cite{Banfi:2010xy,Shao:2026ssz,Jobilal:2026eec}:
\begin{equation}
  \int_{-\infty}^\infty \dd q_\parallel \int \frac{\dd^2\bb}{(2\pi)^2}\, e^{i(q_\parallel b_\parallel + q_\perp b_\perp)}\,\widetilde{\mathcal{W}}(b)
  = \frac{1}{\pi}\int_0^\infty \dd b\,\cos(q_\perp b)\,\widetilde{\mathcal{W}}(b).
  \label{eq:1d_cosine}
\end{equation}
Equation~\eqref{eq:1d_cosine} provides a direct bridge between the measured momentum-space $\Kperp$ distribution and the coordinate-space dipole and Sudakov factors at impact parameter $b$.

\subsection{Fourier-cosine transform and two-scale double ratio}

Motivated by the one-dimensional reduction in eq.~\eqref{eq:1d_cosine}, we define the self-normalized Fourier-cosine moment of the energy-weighted $\Kperp$ spectrum and its nuclear modification ratio at impact parameter $b$:
\begin{equation}
  \mathcal{C}_{pA}^{(n)}(b) \equiv \frac{\int \dd\Kperp\,\cos(\Kperp b)\,\frac{\dd\SigmaHF^{(n),pA}}{\dd\Kperp}}{\int \dd\Kperp\,\frac{\dd\SigmaHF^{(n),pA}}{\dd\Kperp}},
  \qquad \mathcal{R}_{b,\mathrm{fac}}^{\mathrm{HF}}(b) \equiv \frac{\mathcal{C}_{pA}^{(n)}(b)}{\mathcal{C}_{pp}^{(n)}(b)}.
  \label{eq:Rb_def}
\end{equation}
To understand the physical information encoded in eq.~\eqref{eq:Rb_def}, consider first the ideal factorized back-to-back limit in which the folded recoil spectrum ($\Kperp = |q_\perp| \ge 0$) is governed by a single representative hard scale $\mu$ and takes the factorized convolution form
\begin{equation}
  \frac{\dd\Sigma^{pA}}{\dd\Kperp}
  = \frac{2}{\pi}\int_0^\infty\dd b'\,\cos(\Kperp b')\,
     \widetilde{\mathcal{S}}_A(b')\,
     \exp\!\left[-S_P(b';\mu,m_Q) - S_{\mathrm{NP}}(b';\mu,\CNP)\right],
  \label{eq:factorized_recoil}
\end{equation}
where $\widetilde{\mathcal{S}}_A(b)$ denotes the effective coordinate-space target correlator. Taking the inverse Fourier-cosine transform~\eqref{eq:Rb_def} of eq.~\eqref{eq:factorized_recoil} over $0 \le \Kperp < \infty$ converts the momentum-space convolution into a pointwise product in $b$ space. Because $S_P(b;\mu,m_Q)$ and $S_{\mathrm{NP}}(b;\mu,\CNP)$ are properties of the vacuum parton shower and depend only on the hard scale $\mu$ and quark mass $m_Q$, they cancel between numerator and denominator in the nuclear ratio:
\begin{equation}
  \mathcal{R}_{b,\mathrm{fac}}^{\mathrm{HF}}(b) = \frac{\widetilde{\mathcal{S}}_A(b)/\widetilde{\mathcal{S}}_A(0)}{\widetilde{\mathcal{S}}_p(b)/\widetilde{\mathcal{S}}_p(0)},
  \qquad \frac{\partial\mathcal{R}_{b,\mathrm{fac}}^{\mathrm{HF}}}{\partial\CNP} = 0,
  \qquad \frac{\partial\mathcal{R}_{b,\mathrm{fac}}^{\mathrm{HF}}}{\partial S_P} = 0.
  \label{eq:thm4}
\end{equation}
In this ideal factorized limit, any scale-independent collinear nPDF normalization cancels out ($\mathcal{R}_{b,\mathrm{fac}}^{\mathrm{nPDF}}(b) = 1$), whereas the Gaussian broadening baseline~\eqref{eq:gauss_base} reduces to the simple Gaussian ratio $\mathcal{R}_{b,\mathrm{fac}}^{\mathrm{gauss}}(b) = \exp(-\Delta_{\mathrm{Pb}}^2 b^2 / 4)$. In a realistic collider measurement, however, eq.~\eqref{eq:factorized_recoil} holds only approximately: finite $p_T$-window bounds truncate the upper $\Kperp$ integration at $\Kperp^{\max} = \sqrt{2}\,p_{T,\max}$, the hard scale $\mu$ and fragmentation scale $\mu_b$ vary across the $(\xi_1,\xi_2)$ phase space, and the threshold matching factor $g_{\mathrm{match}}$ mixes $W_{\mathrm{soft}}$ and $W_{\mathrm{hard}}$ at large $|\bK'|$. In section~\ref{sec:results_cosine}, we therefore evaluate the finite-acceptance projected cosine ratio $\mathcal{R}_{\cos}^{\mathrm{proj}}(b)$ directly from our full numerical $\Kperp$ spectra over the well-resolved region $\mathcal{C}_{pp}(b) \ge 0.16$, and compare it explicitly against the underlying AAMQS dipole ratio $S_A(b)/S_p(b)$ to quantify where the factorized picture holds and where finite-acceptance effects take over.

As a complementary momentum-space diagnostic that does not require Fourier inversion, we also study the two-scale nuclear double ratio across adjacent transverse-momentum intervals $\mu_1 < \mu_2$~\cite{Lin:2026lz},
\begin{equation}
  \mathcal{D}_{12}(\Kperp) \equiv \frac{\big[\dd\SigmaHF^{p\mathrm{Pb}}/\dd\Kperp \big/ \dd\SigmaHF^{pp}/\dd\Kperp\big]_{\mu_1}}{\big[\dd\SigmaHF^{p\mathrm{Pb}}/\dd\Kperp \big/ \dd\SigmaHF^{pp}/\dd\Kperp\big]_{\mu_2}},
  \label{eq:D12_def}
\end{equation}
which cancels scale-insensitive multiplicative factors while isolating the scale evolution of nuclear broadening from $\mu_1$ to $\mu_2$. Finally, to compare the nonperturbative robustness of different dimensionless nuclear observables $\mathcal{O}$ under the five-point scan $\CNP \in \{1.5, 1.9, 2.3, 2.9, 3.5\}$ relative to the no-broadening baseline $\mathcal{O}_0 = 1$, we define the signal-to-uncertainty ratio
\begin{equation}
  \mathcal{F}_{\mathrm{NP}}(\mathcal{O}) \equiv \frac{|\mathcal{O}(\CNP=2.3) - \mathcal{O}_0|}{\frac{1}{2}[\max_{\CNP}\mathcal{O} - \min_{\CNP}\mathcal{O}]}.
  \label{eq:FOM_def}
\end{equation}
A large value $\mathcal{F}_{\mathrm{NP}}(\mathcal{O}) \gg 1$ indicates that the nuclear modification signal $\mathcal{O} - 1$ is much larger than the half-width of the nonperturbative Sudakov variation band.

\section{Numerical results and phenomenology}
\label{sec:results}

The figures and table in this section are not independent diagnostics; they are organized as a single validation chain that tests each theoretical statement of sections~\ref{sec:framework}--\ref{sec:proj} in turn. We start in section~\ref{sec:lhcb_bench} by confronting the leading-order massive hybrid-CGC and Sudakov calculation with published LHCb $\DD$ data in $\pp$ and $\pPb$ collisions, which validates the away-side description and illustrates how the mirror-antisymmetric HF-ATEEC removes the mid-angle DPS pedestal directly from experimental histograms. Once this data benchmark is fixed, section~\ref{sec:results_kperp} moves to LHC Run~3--4 forward kinematics at $\sqrt{s_{NN}} = 8.16\tev$ ($2.7 < y_{1,2} < 3.7$) to examine how transverse-energy weighting and the orthogonal projection $\Kperp \equiv \aT$ suppress longitudinal fragmentation smearing and separate nonlinear saturation from collinear shadowing and Gaussian broadening across $p_T$ intervals. That comparison reveals a wider baseline separation for $\BB$ than for $\DD$, motivating section~\ref{sec:results_deadcone}, where we isolate the perturbative SCET/bHQET dead-cone mass effect from fragmentation hardness and target-$x_A$ kinematics. Section~\ref{sec:results_cosine} then tests how the momentum-space broadening maps onto the impact-parameter cosine ratio $\mathcal{R}_{\cos}^{\mathrm{proj}}(b)$ and the two-scale double ratio $\mathcal{D}_{12}(\Kperp)$. Finally, section~\ref{sec:results_sys} tracks the broadening continuously across nuclear system size from $\pp$ through light-ion $\pO$ to central $\pPb$ and compares initial saturation-scale variations against matching-scale, $\CNP$, and detector-resolution uncertainties.

Before turning to the physics plots, we summarize the common numerical setup; the analytic $\Kperp$ inversion and change-of-variables Jacobian are recorded in appendix~\ref{app:kperp}. All proton and nuclear targets as well as all $\CNP$ grid points are integrated with shared scrambled Sobol quasi-Monte Carlo sequences so that sampling variance cancels in ratios. The angular HF-TEEC and HF-ATEEC distributions are evaluated on a mirrored 49-node $\delta\phi$ grid with $2^{13}$ points per scramble across eight independent scrambles. For the orthogonal projection $\Kperp$, rather than filling finite-width histograms, we invert $\Kperp(p_{T1}, p_{T2}, \delta\phi)$ analytically at 48--83 fixed knots up to the exact kinematic boundary $\Kperp^{\max} = \sqrt{2}\,p_{T,\max}$, eliminating binning noise from both the spectra and the downstream cosine moments $\mathcal{R}_{\cos}^{\mathrm{proj}}(b)$. The comparison distributions in $|\bq_T|$ and $p_{\mathrm{out}}$, together with the Gaussian detector-resolution checks ($\sigma(p_T)/p_T = 1.5\%$ and $\sigma(\phi) = 10\,\mathrm{mrad}$ per meson), are computed via 11-dimensional phase-space integration with $2^{16}$ points across eight scrambles. Statistical uncertainties on all ratios are obtained by paired leave-one-scramble-out jackknife resampling.

\subsection{\texorpdfstring{Comparison with LHCb forward $\DD$ data}{Comparison with LHCb forward D0 D0bar data}}
\label{sec:lhcb_bench}

\begin{figure}[tbp]
  \centering
  \includegraphics[width=\textwidth]{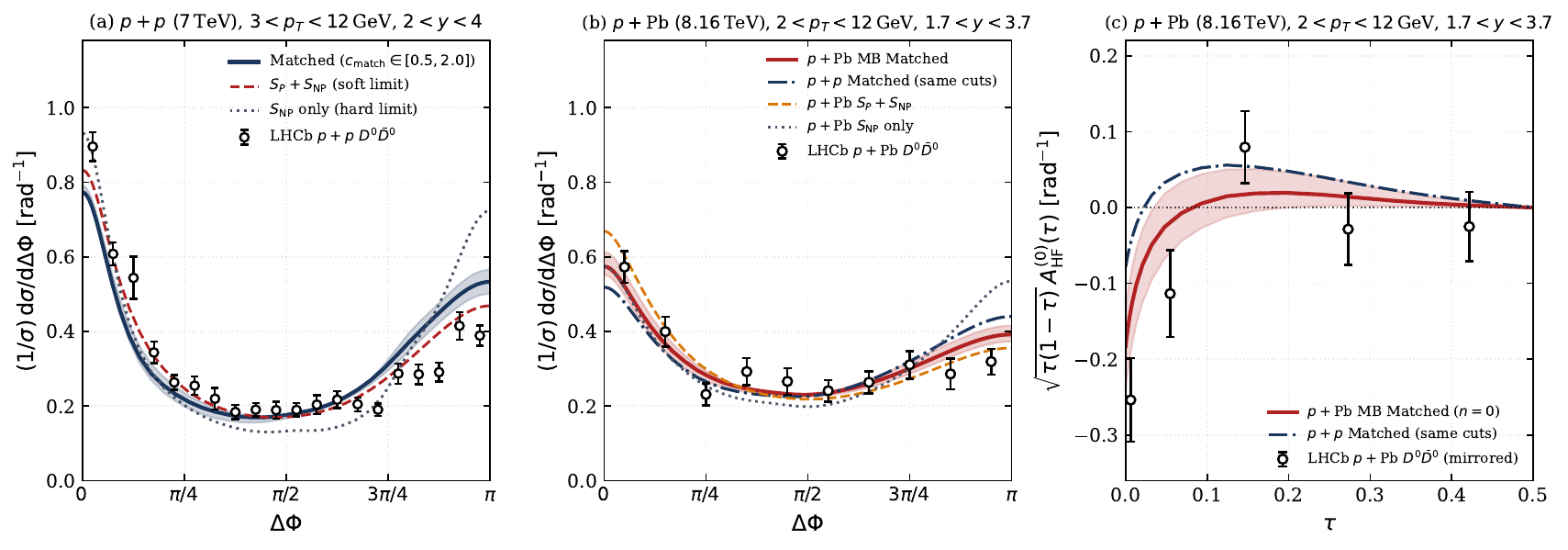}
  \caption{Comparison with LHCb forward $\DD$ azimuthal measurements~\cite{LHCb:2012aiv,LHCb:2020jse}. (a)~$\pp$ at $\sqrt{s}=7\tev$ with the matched CGC+Sudakov band ($c_{\mathrm{match}}\in[0.5,2.0]$) and soft/nonperturbative limits. (b)~Minimum-bias $\pPb$ at $\sqrt{s_{NN}}=8.16\tev$ alongside the matched $\pp$ reference at identical kinematics. (c)~Mirrored HF-ATEEC distribution $\sqrt{\tau(1-\tau)}A_{\mathrm{HF}}^{(0)}(\tau)$ of the LHCb $\pPb$ data compared with $\pPb$ and $\pp$ predictions.}
  \label{fig:lhcb}
\end{figure}

Any prediction for energy-weighted or projected heavy-flavor correlators at LHC Run~3--4 must first rest on a validated description of unweighted azimuthal correlations in existing data. We therefore begin by testing our massive hybrid-CGC and Sudakov setup against the two published LHCb open-charm pair measurements at their exact experimental kinematics: $\pp$ collisions at $\sqrt{s} = 7\tev$ ($3 < p_{T1}, p_{T2} < 12\gev$, $2.0 < y_{1,2} < 4.0$)~\cite{LHCb:2012aiv} and minimum-bias $\pPb$ collisions at $\sqrt{s_{NN}} = 8.16\tev$ ($2 < p_{T1}, p_{T2} < 12\gev$, $1.7 < y_{1,2} < 3.7$)~\cite{LHCb:2020jse}. Because the published LHCb $b\bar{b}$ azimuthal measurement reconstructs inclusive $J/\psi$ pairs from $B$-hadron decays~\cite{LHCb:2017bmm}, we restrict direct experimental comparisons to fully reconstructed prompt $\DD$ pairs. We display the $\pp$ and $\pPb$ datasets in separate panels of figure~\ref{fig:lhcb} so that genuine nuclear modifications are not conflated with the different collision energies, $p_T$ thresholds, and rapidity windows of the two runs.

We first verify the role of threshold-matched Sudakov resummation in the $\pp$ baseline shown in figure~\ref{fig:lhcb}(a). In the back-to-back region $|\Delta\Phi - \pi| \lesssim \pi/3$, the nonperturbative-only curve ($S_{\mathrm{NP}}$, $g_{\mathrm{match}} \equiv 0$) is too narrow and overestimates the peak at $\Delta\Phi = \pi$, whereas the pure soft-resummed curve ($S_P + S_{\mathrm{NP}}$, $g_{\mathrm{match}} \equiv 1$) over-damps the large-angle region $|\Delta\Phi| \lesssim 2\,\mathrm{rad}$. The threshold-matched calculation~\eqref{eq:W_match} with $c_{\mathrm{match}} \in [0.5, 2.0]$ interpolates smoothly between the two regimes and reproduces the measured away-side and mid-angle shape across $\Delta\Phi \in [1.0, \pi]\,\mathrm{rad}$. Near $\Delta\Phi \to 0$, the leading-order $g+p \to c\bar{c}$ channel undershoots the narrow near-side charm peak, where next-to-leading-order $2 \to 3$ partonic channels—specifically final-state gluon splitting $g \to c\bar{c}$ recoiling against a hard jet—provide the dominant contribution~\cite{Salazar:2026ssz}.

Moving from $\pp$ to minimum-bias $\pPb$ collisions at $\sqrt{s_{NN}} = 8.16\tev$ in figure~\ref{fig:lhcb}(b) reveals two distinct modifications when compared with the matched $\pp$ reference evaluated at the same $8.16\tev$ kinematics ($2 < p_T < 12\gev$, $1.7 < y < 3.7$). On the away side ($\Delta\Phi \gtrsim 2.2\,\mathrm{rad}$), coherent multiple scattering off the lead nucleus depletes the back-to-back peak at $\Delta\Phi = \pi$ and broadens the slope. In the intermediate angular region around $\Delta\Phi \simeq \pi/2$, the $\pPb$ data sit above the single-parton-scattering curve due to the nuclear-enhanced DPS pedestal ($F_{\mathrm{DPS}}^{p\mathrm{Pb}} \simeq 3\text{--}4$), which raises the minimum of the azimuthal distribution across all angles.

To remove this uncorrelated DPS pedestal without model-dependent background subtraction, we exploit the fact that the LHCb $\pPb$ histogram is binned symmetrically about $\Delta\Phi = \pi/2$ and construct the odd HF-ATEEC combination $\sqrt{\tau(1-\tau)}A_{\mathrm{HF}}^{(0)}(\tau) = y(\pi - \delta\phi) - y(\delta\phi)$ in figure~\ref{fig:lhcb}(c). By virtue of eq.~\eqref{eq:thm2b}, any constant DPS offset and all even harmonics cancel identically in the reflected difference. At very small $\tau \lesssim 0.08$, the experimental points dip negative because the narrow near-side NLO $g \to c\bar{c}$ peak at $\Delta\Phi \to 0$ ($\tau \to 1$) exceeds the away-side peak at $\Delta\Phi \to \pi$ ($\tau \to 0$). Outside this near-collinear endpoint, the matched $\pPb$ prediction reproduces the measured odd angular profile across $\tau \in [0.10, 0.50]$ and shows a clear suppression relative to the $\pp$ reference at $\tau \lesssim 0.20$ driven by nuclear recoil broadening.

\subsection{\texorpdfstring{HF-TEEC nuclear modification and single-axis $\Kperp \equiv \aT$ projections}{HF-TEEC nuclear modification and single-axis Kperp equals aT projections}}
\label{sec:results_kperp}

\begin{figure}[tbp]
  \centering
  \includegraphics[width=\textwidth]{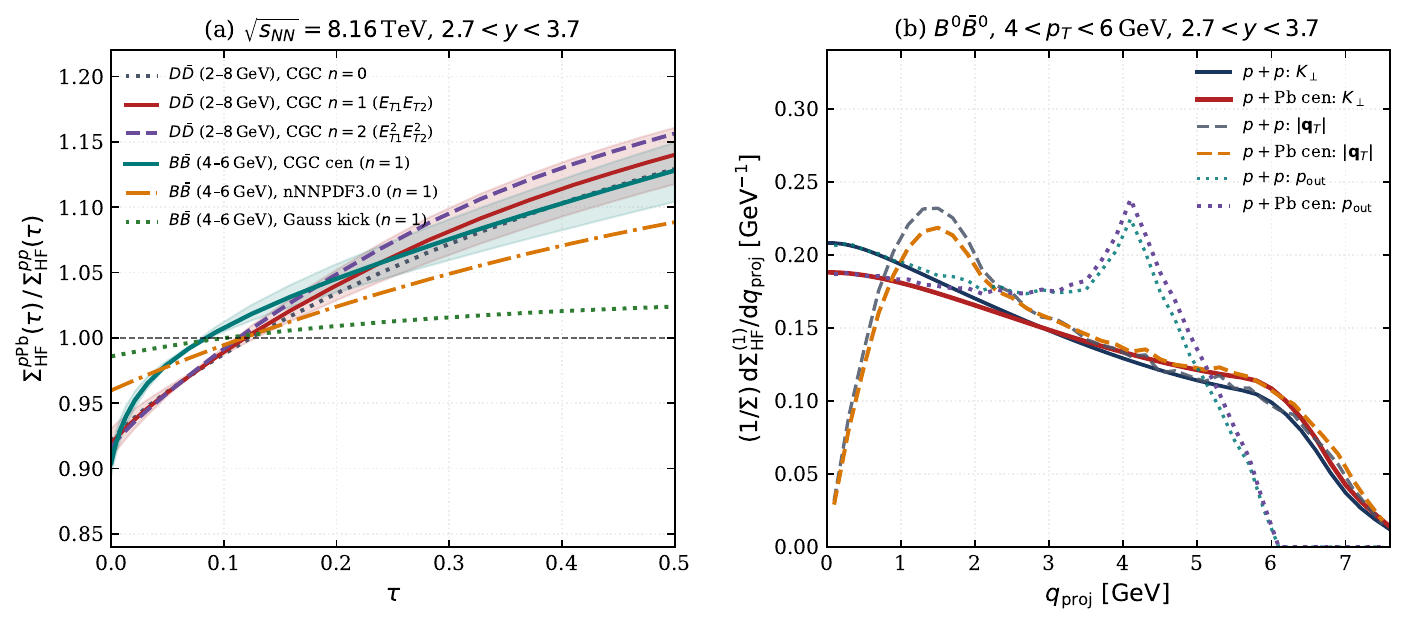}
  \caption{(a)~Away-side HF-TEEC nuclear ratio $\SigmaHF^{p\mathrm{Pb}}(\tau)/\SigmaHF^{pp}(\tau)$ comparing energy weights $n=0,1,2$ in wide-window $\DD$ and contrasting central CGC saturation in $\BB$ against \texttt{nNNPDF3.0} shadowing and Gaussian broadening; shaded bands show the five-point $\CNP$ range. (b)~Self-normalized transverse-momentum projections $q_{\mathrm{proj}} \in \{\Kperp, |\bq_T|, p_{\mathrm{out}}\}$ for $\BB$ ($4<p_T<6\gev$, $n=1$) in $\pp$ and central $\pPb$ collisions.}
  \label{fig:teec_proj}
\end{figure}

Having verified that the massive hybrid-CGC framework describes the away-side angular structure of existing LHCb charm data, we now turn to forward LHC Run~3--4 kinematics at $\sqrt{s_{NN}} = 8.16\tev$ ($2.7 < y_{1,2} < 3.7$) to examine how transverse-energy weighting and single-axis momentum projection isolate the saturation scale $\Qs(x_A)$. Figure~\ref{fig:teec_proj}(a) presents the self-normalized away-side HF-TEEC nuclear modification ratio $\SigmaHF^{p\mathrm{Pb}}(\tau)/\SigmaHF^{pp}(\tau)$ over $\tau \in [0, 0.5]$. In the central-CGC calculation, coherent multiple scattering off the dense nuclear color field depletes the collinear back-to-back peak ($\tau \to 0$, where the ratio drops to $0.82\text{--}0.85$) and redistributes angular weight toward larger acoplanarity ($\tau \in [0.15, 0.50]$, where the ratio rises above $1.20$). In the wide charm acceptance $2 < p_T < 8\gev$, increasing the transverse-energy weight from $n = 0$ to $n = 1$ and $n = 2$ progressively steepens this away-side tilt by emphasizing harder meson pairs from larger parent-quark momentum fractions $\xi$. At the same time, evaluating the DPS moment ratio~\eqref{eq:thm2a} with our single-inclusive and SPS pair spectra shows that energy weighting suppresses the relative DPS pedestal in wide-window $\DD[2,8]\gev$ to $f_{\mathrm{DPS}}^{(1)}/f_{\mathrm{DPS}}^{(0)} = 0.854$ ($0.850$ in minimum-bias $\pPb$) for $n = 1$ and $0.702$ ($0.694$) for $n = 2$. In wide-window bottom production $\BB[3,9]\gev$, the corresponding DPS ratios are $0.943$ ($n=1$) and $0.885$ ($n=2$), while inside narrow $2\gev$-wide bins such as $\DD[2,4]\gev$ ($0.965$ for $n=1$) and $\BB[4,6]\gev$ ($0.992$ for $n=1$), the variation of $E_T$ across the bin is small and DPS suppression relies instead on the exact mirror cancellation of HF-ATEEC~\eqref{eq:thm2b}. For $\BB$ pairs in $4 < p_T < 6\gev$, figure~\ref{fig:teec_proj}(a) also demonstrates that the central-CGC HF-TEEC ratio exhibits a much stronger angular redistribution than either collinear \texttt{nNNPDF3.0} shadowing or linear Gaussian broadening, both of which remain nearly flat across $\tau \in [0, 0.5]$ with narrow $\CNP$ variation bands.

While the angular HF-TEEC ratio in figure~\ref{fig:teec_proj}(a) clearly displays nuclear broadening in $\tau$, it integrates over the absolute transverse momenta of the mesons within the bin. To measure the nuclear momentum transfer directly in GeV units without contamination from longitudinal fragmentation recoil, we compare in figure~\ref{fig:teec_proj}(b) the three candidate transverse-momentum variables $q_{\mathrm{proj}} \in \{\Kperp, |\bq_T|, p_{\mathrm{out}}\}$ for $\BB$ pairs ($4 < p_T < 6\gev$, $n=1$) in $\pp$ and central $\pPb$ collisions. The numerical curves confirm the small-angle expansion in eqs.~\eqref{eq:qpar_noise}--\eqref{eq:Kperp_noise}. Because the two-dimensional vector imbalance $|\bq_T| = \sqrt{q_\parallel^2 + \Kperp^2}$ contains the linear longitudinal fragmentation recoil $q_\parallel = 2\bar{\xi}k_T\epsilon + \mathcal{O}(\delta\phi^2)$, its differential distribution is forced to vanish at $|\bq_T| = 0$ by the radial phase-space factor and peaks around $|\bq_T| \simeq 1.5\gev$, masking the region $|\bK'| \lesssim \Qs$ where saturation effects originate. The trigger-projected variable $p_{\mathrm{out}} = p_{T2}|\sin\Delta\Phi|$ peaks at zero but suffers from linear $\mathcal{O}(\epsilon)$ smearing and develops a sharp kinematic Jacobian cusp at the lower bin boundary $p_{T,\min} = 4\gev$ before vanishing at $p_{T,\max} = 6\gev$. By contrast, the orthogonal projection $\Kperp \equiv \aT$ eliminates linear fragmentation fluctuations ($\partial\Kperp/\partial\epsilon|_{\epsilon=0} = 0$), peaks smoothly at $\Kperp = 0$, and decays monotonically across the away-side phase space, making the depletion at $\Kperp \lesssim 1.2\gev$ and broadening at $\Kperp \gtrsim 1.5\gev$ in central $\pPb$ collisions directly visible.

\begin{figure}[tbp]
  \centering
  \includegraphics[width=\textwidth]{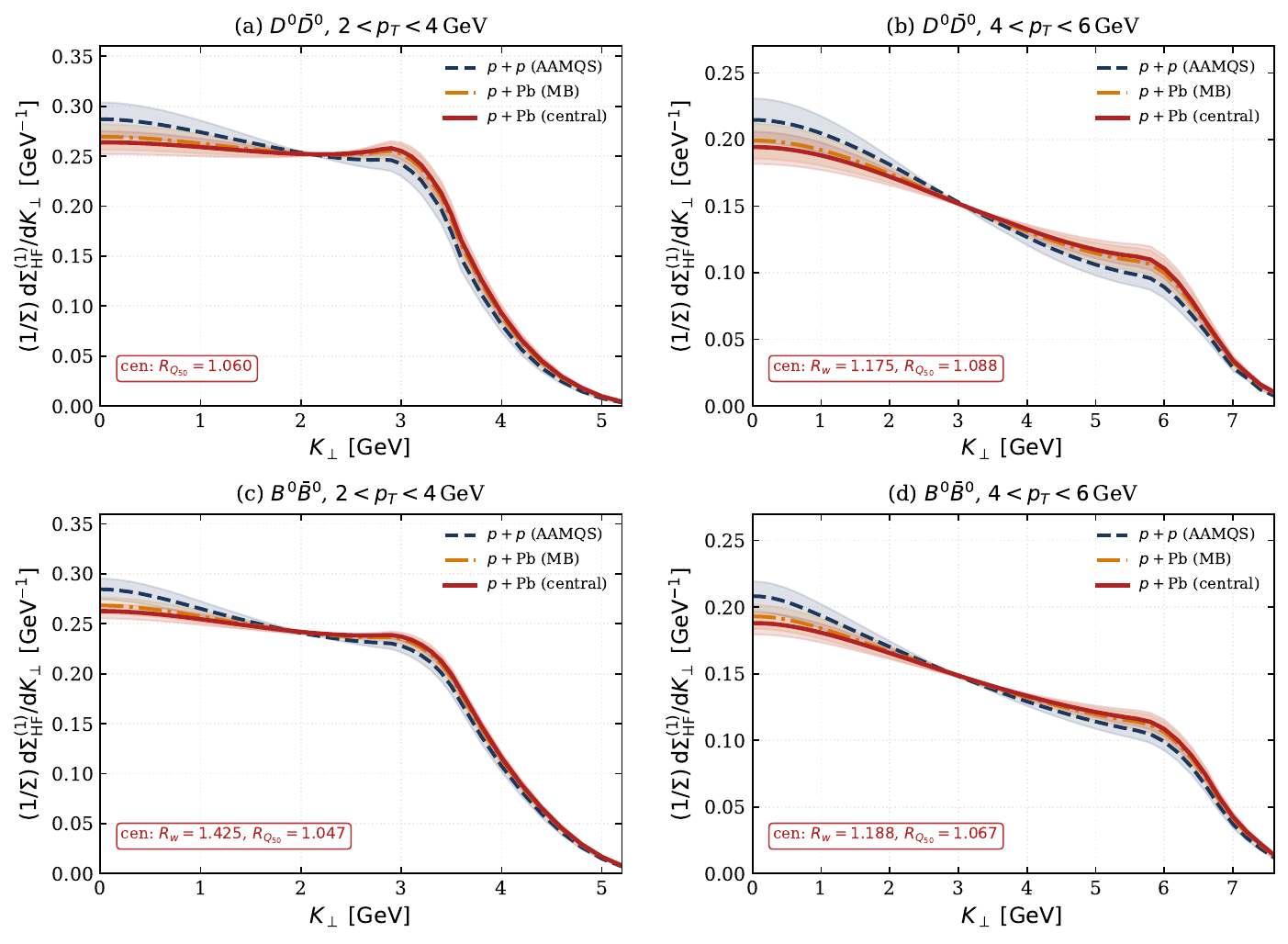}
  \caption{Smooth $\Kperp \equiv \aT$ distributions ($n=1$) in $\pp$, minimum-bias $\pPb$, and central $\pPb$ collisions at $\sqrt{s_{NN}}=8.16\tev$ ($2.7<y_{1,2}<3.7$) for $\DD$ and $\BB$ in $[2,4]\gev$ and $[4,6]\gev$. Shaded bands denote the five-point $\CNP\in[1.5,3.5]$ scan around $\CNP=2.3$.}
  \label{fig:kp_grid}
\end{figure}

Having identified $\Kperp \equiv \aT$ as the projection free of linear fragmentation recoil and kinematic cusps, we map its self-normalized differential spectrum ($n=1$) across heavy-quark flavors ($\DD$ and $\BB$), transverse-momentum windows ($[2,4]\gev$ and $[4,6]\gev$), and nuclear centralities ($\pp$, minimum-bias $\pPb$, and central $\pPb$) in figure~\ref{fig:kp_grid}. Across all four panels, increasing the nuclear target thickness systematically depletes the back-to-back peak at $\Kperp \to 0$ and shifts probability weight into the semi-hard tail $\Kperp \sim 1.5\text{--}4\gev$. Comparing the left column ($[2,4]\gev$) with the right column ($[4,6]\gev$) shows that nuclear broadening is strongest at lower meson $p_T$, where the target is probed at smaller longitudinal momentum fractions ($\langle x_A\rangle = 4.46\times 10^{-5}$ for $\DD[2,4]$ and $4.92\times 10^{-5}$ for $\BB[2,4]$, corresponding to $Y_{\mathrm{eff}} = 5.41$ and $5.31$) and where the saturation scale $\Qs(x_A)$ represents a larger fraction of the pair hard scale $\mu$. In particular, figure~\ref{fig:kp_grid}(a) illustrates the onset of a saturated regime in central $\pPb$ collisions for $\DD[2,4]\gev$: coherent multiple scattering flattens the small-$\Kperp$ peak into a broad plateau extending out to $\Kperp \simeq 2\gev$. Because this flattened profile is no longer quasi-Gaussian near the origin, fitting a local Gaussian curvature to $\ln(\dd\Sigma/\dd\Kperp)$ in $\DD[2,4]\gev$ yields a large nominal width ratio $R_w(\Kperp) = 1.731$ with a wide $\CNP$ sensitivity span $[1.519, 4.006]$ ($\mathcal{F}_{\mathrm{NP}} = 0.59$). By contrast, the non-parametric median quantile ratio $R_{Q_{50}} = Q_{50}^{p\mathrm{Pb}} / Q_{50}^{pp}$ remains well-conditioned in this strong-broadening window, giving $R_{Q_{50}} = 1.060$ $[1.047, 1.075]$ in central $\pPb$ ($1.045$ $[1.035, 1.056]$ in minimum-bias $\pPb$) for $\DD[2,4]\gev$ and $R_{Q_{50}} = 1.047$ $[1.039, 1.057]$ ($1.035$ $[1.028, 1.042]$ in minimum-bias $\pPb$) for $\BB[2,4]\gev$, well above the corresponding \texttt{nNNPDF3.0} ($1.033$ and $1.021$) and Gaussian-kick ($1.016$ and $1.009$) baselines.

\begin{table}[tbp]
  \centering
  \small
  \renewcommand{\arraystretch}{1.15}
  \begin{tabular}{@{}lcccc@{}}
    \toprule
    Channel ($p_T$ [GeV]) & CGC central ($\CNP$ range) & CGC MB & \texttt{nNNPDF3.0} & Gaussian kick \\
    \midrule
    $\BB\ [4,6]$ & $\mathbf{1.188}$ $[1.187,\,1.192]$ & $1.129$ & $1.060$ & $1.023$ \\
    $\DD\ [4,6]$ & $\mathbf{1.175}$ $[1.174,\,1.179]$ & $1.121$ & $1.097$ & $1.034$ \\
    $\BB\ [6,8]$ & $1.143$ $[1.141,\,1.147]$ & $1.101$ & $1.057$ & $1.018$ \\
    $\DD\ [6,8]$ & $1.131$ $[1.128,\,1.136]$ & $1.093$ & $1.079$ & $1.022$ \\
    \bottomrule
  \end{tabular}
  \caption{Representative nuclear Gaussian width ratios $R_w(\Kperp, n=1)$ at $\sqrt{s_{NN}}=8.16\tev$ ($2.7<y_{1,2}<3.7$, baseline $\CNP=2.3$). Brackets show the min--max span across $\CNP\in\{1.5,1.9,2.3,2.9,3.5\}$; paired jackknife Monte Carlo errors are $0.002$--$0.006$ for central CGC and smaller for minimum-bias and non-saturation baselines.}
  \label{tab:master}
\end{table}

In the intermediate $[4,6]\gev$ and higher $[6,8]\gev$ windows, where the away-side peak remains well-behaved and quasi-Gaussian over $\Kperp \le k_{\mathrm{fit}}$, the Gaussian width ratio $R_w(\Kperp, n=1)$ provides a precise quantitative summary of the broadening hierarchy, as compiled in table~\ref{tab:master}. Three physical features stand out in table~\ref{tab:master}. First, in the $[4,6]\gev$ window, central CGC saturation produces an $18.8\%$ width broadening ($R_w = 1.188 \pm 0.002$) for $\BB$ and a $17.5\%$ broadening ($R_w = 1.175 \pm 0.006$) for $\DD$, with minimum-bias $\pPb$ giving $1.129$ and $1.121$, respectively. Across the entire five-point scan $\CNP \in [1.5, 3.5]$, the central CGC width ratio varies by only $0.004\text{--}0.005$ (corresponding to nonperturbative stability figures of merit $\mathcal{F}_{\mathrm{NP}} = 93.4$ for $\BB[4,6]$ and $72.2$ for $\DD[4,6]$, and $\mathcal{F}_{\mathrm{NP}} = 132.8$ for minimum-bias $\BB[4,6]$). Second, both non-saturation baselines fall well below the CGC prediction: collinear \texttt{nNNPDF3.0} shadowing yields $R_w = 1.060$ for $\BB[4,6]$ and $1.097$ for $\DD[4,6]$, while the linear Gaussian kick~\eqref{eq:gauss_base} yields only $1.023$ and $1.034$. The large gap between the full rcBK calculation ($1.188$) and the Gaussian kick ($1.023$)—despite sharing the same saturation-scale shift $\Delta_{\mathrm{Pb}}^2(x_A)$—demonstrates that heavy-meson pair broadening at $p_T \sim 4\text{--}6\gev$ is driven primarily by the power-law semi-hard tail ($l \sim 2\text{--}6\gev$) of the nonlinear rcBK dipole rather than by soft Gaussian diffusion at $l \lesssim 1\gev$. Third, stepping up to the $[6,8]\gev$ window reduces the central CGC broadening to $R_w = 1.143$ ($\mathcal{F}_{\mathrm{NP}} = 51.5$) for $\BB$ and $1.131$ ($\mathcal{F}_{\mathrm{NP}} = 32.9$) for $\DD$ as $\langle x_A\rangle$ increases to $9.52\times 10^{-5}$ ($Y_{\mathrm{eff}} = 4.65$) and $1.15\times 10^{-4}$ ($Y_{\mathrm{eff}} = 4.47$) and $\Qs^2/\mu^2$ decreases. By contrast, the \texttt{nNNPDF3.0} baseline decreases much more slowly with $p_T$ ($1.060 \to 1.057$ for $\BB$ and $1.097 \to 1.079$ for $\DD$), providing a distinct scale-evolution signature that we exploit via the two-scale double ratio in section~\ref{sec:results_cosine}.

\subsection{Heavy-quark mass hierarchy and mechanism decomposition}
\label{sec:results_deadcone}

\begin{figure}[tbp]
  \centering
  \includegraphics[width=\textwidth]{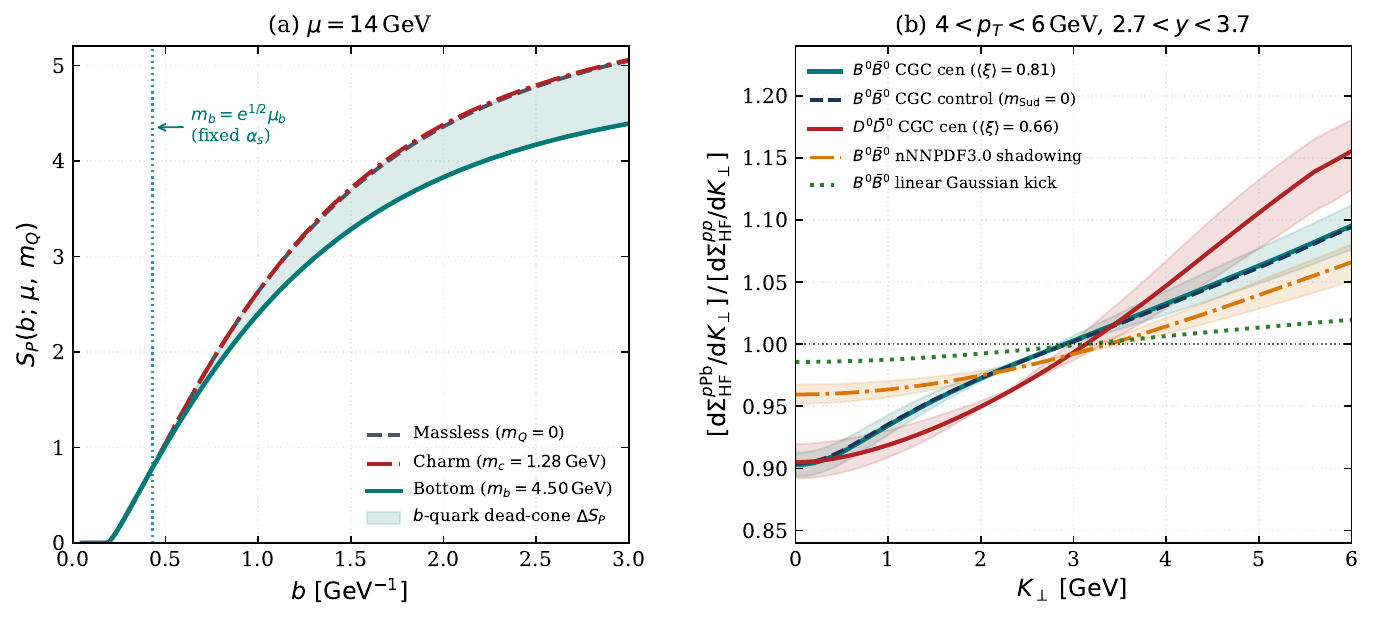}
  \caption{(a)~Perturbative SCET/bHQET Sudakov exponent $S_P(b; \mu=14\gev, m_Q)$ for $m_Q=0$, $m_c$, and $m_b$, marking the fixed-coupling crossover $m_b = e^{1/2}\mu_b$. (b)~Self-normalized nuclear ratio $[\dd\SigmaHF^{p\mathrm{Pb}}/\dd\Kperp]/[\dd\SigmaHF^{pp}/\dd\Kperp]$ at $p_T\in[4,6]\gev$ comparing $\BB$ CGC, the massless-Sudakov benchmark ($m_{\mathrm{Sud}}=0$), $\DD$ CGC, \texttt{nNNPDF3.0}, and Gaussian broadening; shaded bands show the five-point $\CNP$ range.}
  \label{fig:deadcone}
\end{figure}

Table~\ref{tab:master} raises an important physical question: at matched meson transverse momentum $4 < p_T < 6\gev$, why do bottom-meson pairs exhibit both a slightly larger CGC broadening ($R_w = 1.188$ vs.\ $1.175$) and a much wider separation from the collinear \texttt{nNNPDF3.0} baseline ($R_w^{\mathrm{CGC}} - R_w^{\mathrm{nPDF}} = 0.127$ for $\BB$ vs.\ $0.078$ for $\DD$) than charm-meson pairs? Because the heavy-quark mass $m_Q$ enters our framework in three distinct places—the perturbative SCET/bHQET Sudakov exponent $S_P(b; \mu, m_Q)$, the fragmentation function $D_{H/Q}(\xi, \mu)$ and finite-mass map~\eqref{eq:frag_map}, and the massive hard matrix element $\mathcal{H}_{gA}$—we perform a controlled decomposition in figure~\ref{fig:deadcone} to identify which mechanism drives the flavor hierarchy.

We first examine the impact-parameter profile of the perturbative SCET/bHQET Sudakov exponent $S_P(b; \mu=14\gev, m_Q)$ in figure~\ref{fig:deadcone}(a) for $m_Q = 0$, $m_c = 1.28\gev$, and $m_b = 4.50\gev$. In exact agreement with our analytical threshold condition~\eqref{eq:deadcone_onset}, the charm mass sits close to the fixed-coupling crossover scale $e^{1/2}\mu_b^{\min} \simeq 1.23\gev$, so $S_P(b; \mu, m_c)$ is nearly indistinguishable from the massless exponent $S_P(b; \mu, 0)$ over the entire impact-parameter range (e.g.\ $S_P = 4.253$ vs.\ $4.260$ at $b = 1.5\gev^{-1}$). For bottom quarks, by contrast, $m_b$ exceeds $e^{1/2}\mu_b$ once $b \gtrsim 0.43\gev^{-1}$, where the bHQET dead-cone screening term $-2C_F\ln(m_b^2/\bar{\mu}^2)$ freezes the final-state double logarithm and lowers $S_P(b; \mu, m_b)$ to $3.619$ at $b = 1.5\gev^{-1}$. In the individual $\pp$ and $\pPb$ spectra, this dead-cone suppression of $S_P$ reduces vacuum radiative broadening and sharpens the back-to-back $\BB$ peak relative to a massless parton at the same hard scale $\mu$.

To test whether this perturbative Sudakov dead-cone term is responsible for the enhanced nuclear ratio of $\BB$ relative to $\DD$, figure~\ref{fig:deadcone}(b) compares the full massive $\Kperp$ nuclear modification ratio $[\dd\SigmaHF^{p\mathrm{Pb}}/\dd\Kperp]/[\dd\SigmaHF^{pp}/\dd\Kperp]$ in $[4,6]\gev$ against a control calculation ($m_{\mathrm{Sud}} = 0$, open green markers) in which we set $m_Q \to 0$ solely inside the perturbative Sudakov exponent $S_P(b; \mu, m_Q)$ while retaining the physical mass $m_b = 4.50\gev$ in the hard matrix element $\mathcal{H}_{gA}$ and the fragmentation map~\eqref{eq:frag_map}. Remarkably, turning off the dead-cone mass term in $S_P$ leaves the self-normalized nuclear ratio in figure~\ref{fig:deadcone}(b) almost completely unchanged: the extracted width ratio shifts by less than $0.1\%$, from $R_w(\Kperp) = 1.1875$ to $1.1868$ ($R_{Q_{50}}$ from $1.0669$ to $1.0656$), and for $\DD[4,6]\gev$ it is identical to five significant digits ($1.1747$). The physical reason follows directly from the factorized structure of section~\ref{sec:proj}: because $S_P(b; \mu, m_Q)$ multiplies both the proton and nuclear integrands in $b$ space, its leading effect cancels in self-normalized $\pPb/\pp$ ratios. Consequently, the superior baseline separation of $\BB$ over $\DD$ in table~\ref{tab:master} and figure~\ref{fig:deadcone}(b) is governed by three non-Sudakov mass effects: (i) the much harder $b \to B^0$ fragmentation ($\langle\bar{\xi}\rangle_{n=1} = 0.806$ in $\BB[4,6]$ vs.\ $0.664$ in $\DD[4,6]$), which transmits the nuclear transverse kick $|\bK'|$ to the mesons with minimal $\mathcal{O}(\epsilon^2)$ dilution; (ii) the smaller parent quark transverse momentum $k_T \simeq p_T/\bar{\xi}$ required at fixed meson $p_T$, which keeps the target longitudinal fraction small ($\langle x_A\rangle = 7.04\times 10^{-5}$, $Y_{\mathrm{eff}} = 4.96$ in $\BB[4,6]$ vs.\ $\langle x_A\rangle = 7.97\times 10^{-5}$, $Y_{\mathrm{eff}} = 4.83$ in $\DD[4,6]$); and (iii) the weaker scale variation of the collinear \texttt{nNNPDF3.0} ratio $R_g^{\mathrm{Pb}}(x_A, \mu)$ at the higher hard scale $\mu \sim 2m_{Tb}$ of the bottom pair, which suppresses spurious collinear-shadowing broadening in $\BB$ ($R_w^{\mathrm{nPDF}} = 1.060$ vs.\ $1.097$ in $\DD$).

\subsection{Projected cosine moments and two-scale double ratios}
\label{sec:results_cosine}

\begin{figure}[tbp]
  \centering
  \includegraphics[width=\textwidth]{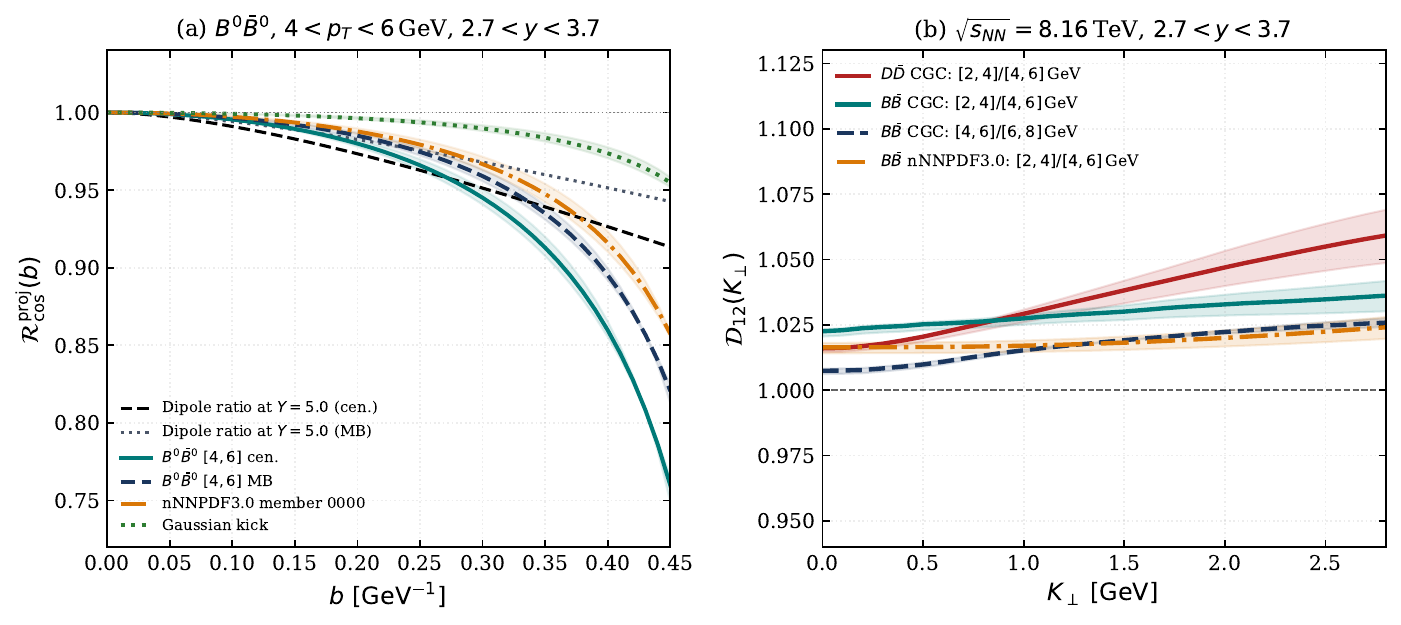}
  \caption{(a)~Finite-acceptance cosine ratios $\mathcal{R}_{\cos}^{\mathrm{proj}}(b)$ from the smooth $\Kperp$ spectra (shown where $\mathcal{C}_{pp}(b)\ge 0.16$) compared with AAMQS dipole ratios. (b)~Two-scale nuclear double ratios $\mathcal{D}_{12}(\Kperp)$ across adjacent $p_T$ windows over their shared support $\Kperp \le 2.8\gev$. Shaded bands in both panels show the five-point $\CNP$ range.}
  \label{fig:deconv}
\end{figure}

The $m_{\mathrm{Sud}} = 0$ control test in section~\ref{sec:results_deadcone} showed that common Sudakov factors largely cancel in self-normalized momentum-space ratios, which explains the narrow $\CNP$ bands found in table~\ref{tab:master}. In section~\ref{sec:proj} we proved that in the idealized factorized back-to-back limit~\eqref{eq:factorized_recoil}, taking the one-dimensional Fourier-cosine transform $\mathcal{R}_{b,\mathrm{fac}}^{\mathrm{HF}}(b)$ cancels common perturbative and nonperturbative Sudakov factors point-by-point in conjugate impact-parameter space $b$. To determine how accurately this relation holds in a realistic finite-acceptance measurement, figure~\ref{fig:deconv}(a) compares the projected cosine ratios $\mathcal{R}_{\cos}^{\mathrm{proj}}(b)$ integrated directly from our smooth $\Kperp$ spectra (over the well-conditioned domain where the denominator $\mathcal{C}_{pp}(b) \ge 0.16$) against the exact underlying AAMQS fundamental dipole ratios $S_A(Y_{\mathrm{eff}}, b)/S_p(Y_{\mathrm{eff}}, b)$ at the effective rapidity $Y_{\mathrm{eff}} = 4.96$ of the $\BB[4,6]\gev$ window.

In the small- and moderate-$b$ region ($b \lesssim 0.30\gev^{-1}$, where $\mathcal{C}_{pp}(b=0.30\gev^{-1}) = 0.532$ is well above zero), the projected cosine ratios $\mathcal{R}_{\cos}^{\mathrm{proj}}(b)$ in $\BB[4,6]\gev$ reproduce both the quadratic $b^2$ falloff and the centrality ordering of the exact AAMQS dipole ratios remarkably closely: at the reference point $b_* = 0.30\gev^{-1}$, the projected calculation gives $\mathcal{R}_{\cos}^{\mathrm{proj}}(b_*) = 0.945$ $[0.941, 0.951]$ for central $\pPb$ and $0.959$ $[0.956, 0.963]$ for minimum-bias $\pPb$, compared with the exact AAMQS single-dipole ratios $S_{\mathrm{Pb}}^{\mathrm{cen}}(b_*)/S_p(b_*) = 0.951$ and $S_{\mathrm{Pb}}^{\mathrm{MB}}(b_*)/S_p(b_*) = 0.968$. The Gaussian-kick baseline gives $\mathcal{R}_{\cos}^{\mathrm{proj}}(b_*) = 0.990$, while the \texttt{nNNPDF3.0} curve gives $0.967$. At larger impact parameters $b \in [0.35, 0.60]\gev^{-1}$ (conjugate to soft transverse momenta $\Kperp \sim 1/b \lesssim 2\gev$), the decaying denominator $\mathcal{C}_{pp}(b)$ and the upper integration cutoff $\Kperp^{\max} = \sqrt{2}\,p_{T,\max}$ amplify finite-acceptance departures from the idealized infinite-domain identity~\eqref{eq:thm4}. In particular, because collinear \texttt{nNNPDF3.0} shadowing modifies the $\Kperp$ shape slightly through the scale dependence of $R_g^{\mathrm{Pb}}(x_A, \mu_b)$, its finite-acceptance cosine ratio also deviates from unity at large $b$, and larger kinematic shifts appear in the higher $[6,8]\gev$ bin where $\mathcal{C}_{pp}(b_*)$ drops to $0.337\text{--}0.414$. Thus, $\mathcal{R}_{\cos}^{\mathrm{proj}}(b)$ at $b \lesssim 0.30\gev^{-1}$ provides a useful, nearly $\CNP$-independent impact-parameter diagnostic, but must be interpreted together with the full finite-acceptance forward calculation rather than equated naively to a bare dipole ratio at large $b$.

Complementing the impact-parameter cosine transform, figure~\ref{fig:deconv}(b) displays the two-scale momentum-space double ratio $\mathcal{D}_{12}(\Kperp)$ defined in eq.~\eqref{eq:D12_def} across adjacent $p_T$ intervals over their common away-side support $\Kperp \le 2.8\gev$. Because overall target normalizations and scale-insensitive systematical factors cancel between the $\mu_1$ numerator and $\mu_2$ denominator, $\mathcal{D}_{12}(\Kperp)$ directly measures how the nuclear broadening evolves with the hard scale. For both $\BB([2,4]/[4,6]\gev)$, $\BB([4,6]/[6,8]\gev)$, and $\DD([4,6]/[6,8]\gev)$, the central-CGC double ratio starts below unity at $\Kperp \to 0$ (reflecting the stronger peak depletion in the lower-$p_T$ bin) and rises monotonically above unity for $\Kperp \gtrsim 1.3\gev$ (reflecting the stronger semi-hard tail enhancement at smaller $x_A$ and larger $\Qs/\mu$). By contrast, the collinear \texttt{nNNPDF3.0} double ratio remains nearly flat and close to unity across the entire $\Kperp$ range, with very narrow $\CNP$ uncertainty bands on all curves. Measuring $\mathcal{D}_{12}(\Kperp)$ across adjacent $p_T$ bins in LHCb Run~3--4 data therefore offers a direct test of the $\Qs(x_A)/\mu$ scale hierarchy of nonlinear gluon saturation.

\subsection{\texorpdfstring{Nuclear-opacity scaling, $\CNP$ dependence, and detector resolution}{Nuclear-opacity scaling, CNP dependence, and detector resolution}}
\label{sec:results_sys}

\begin{figure}[tbp]
  \centering
  \includegraphics[width=\textwidth]{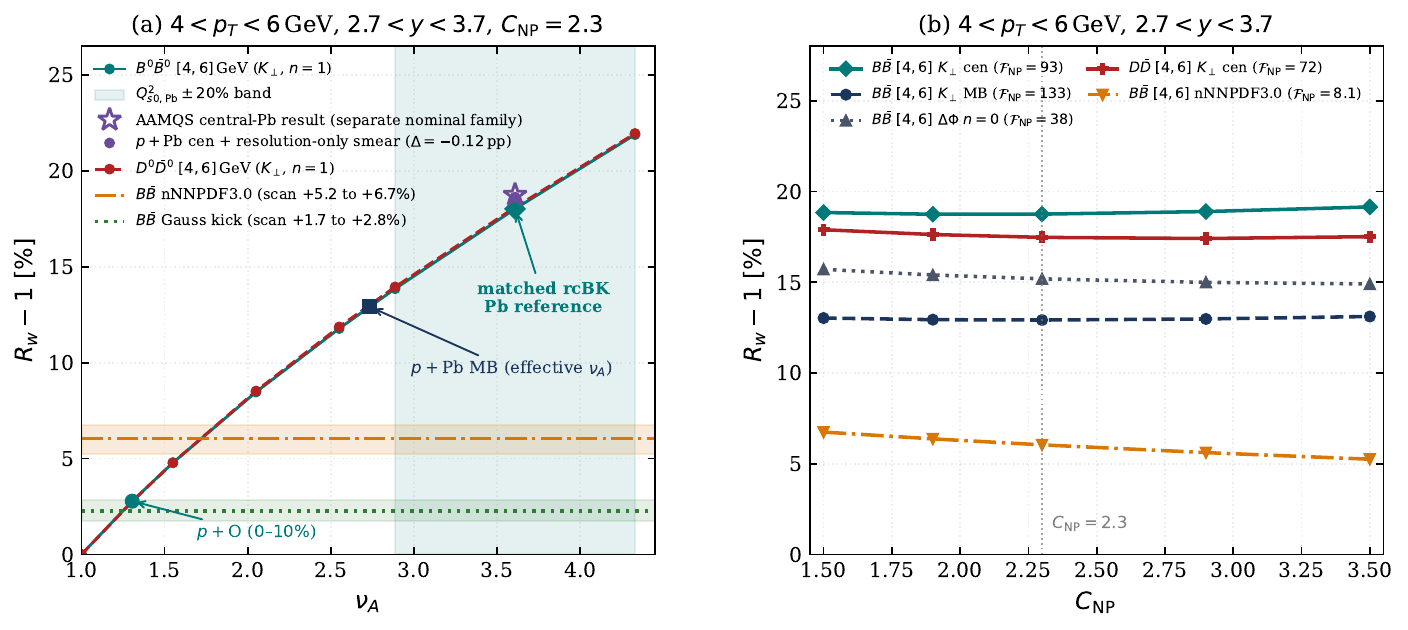}
  \caption{(a)~Matched 1D rcBK opacity scan of $R_w(\Kperp, n=1)$ in $[4,6]\gev$ from $\pp$ through central $\pO$ and minimum-bias $\pPb$ to central $\pPb$, compared with the AAMQS central-Pb reference, the $Q_{s0,\mathrm{Pb}}^2 \pm 20\%$ band, and non-saturation baselines. (b)~Five-point $\CNP$ scan illustrating the stability of $R_w$ and the nonperturbative sensitivity ratio $\mathcal{F}_{\mathrm{NP}}$.}
  \label{fig:sys}
\end{figure}

In the preceding sections, we compared $\pp$ collisions with minimum-bias and central $\pPb$ collisions at fixed initial dipole normalizations. For experimental saturation searches across the LHC Run~3 and Run~4 program—which includes both the $\pO$ light-ion run and high-statistics $\pPb$ data—two final questions must be addressed: how does the $\Kperp$ broadening scale continuously with the nuclear saturation scale $Q_{s0,A}^2$, and can theoretical or experimental systematic uncertainties blur the separation between CGC saturation and the non-saturation baselines?

To answer the first question, figure~\ref{fig:sys}(a) tracks the Gaussian width ratio $R_w(\Kperp, n=1)$ in the $[4,6]\gev$ window as a continuous function of the initial nuclear opacity ratio $\nu_A = Q_{s0,A}^2 / Q_{s0,p}^2 \in [1.0, 4.33]$ obtained by solving the one-dimensional Balitsky rcBK equation. As $\nu_A$ increases from the proton reference ($\nu_A = 1$, $R_w \equiv 1$), the width ratio grows monotonically and approximately linearly with the nuclear saturation scale for both $\BB$ and $\DD$. At the $0\text{--}10\%$ central $\pO$ opacity point ($\nu_{\mathrm{O},0\text{--}10\%} = 1.3054$), the matched rcBK calculation predicts a modest $2.8\%$ width broadening for both bottom and charm pairs ($R_w(\Kperp) = 1.028$, and $1.038$ for $\BB[2,4]\gev$, $1.044$ for $\DD[2,4]\gev$), comparable in magnitude to the linear Gaussian-kick baseline in $\pPb$. This indicates that central $\pO$ collisions serve as an intermediate light-ion control point where linear multiple-scattering approximations begin to transition toward nonlinear saturation. At the minimum-bias $\pPb$ opacity point, the matched rcBK curve rises to $R_w = 1.133$ for $\BB[4,6]$ and $1.112$ for $\DD[4,6]$ (close to the impact-parameter-averaged AAMQS minimum-bias values $1.129$ and $1.121$), and reaches $R_w = 1.180$ ($\BB[4,6]$) and $1.181$ ($\DD[4,6]$) at the matched central-Pb reference $\nu_{\mathrm{Pb},0} = 3.608$ (compared with the AAMQS central-Pb points $1.188$ and $1.175$, shown as open markers in figure~\ref{fig:sys}(a) to distinguish genuine nuclear-opacity scaling from dipole-parametrization differences).

Figure~\ref{fig:sys} also quantifies the complete hierarchy of theoretical and experimental systematic uncertainties in the $[4,6]\gev$ window. The dominant theoretical uncertainty stems from the initial nuclear saturation scale $Q_{s0,\mathrm{Pb}}^2$: varying $\nu_A$ by $\pm 20\%$ around $\nu_{\mathrm{Pb},0} = 3.608$ ($\nu_A \in [2.886, 4.330]$, vertical gray shaded band in figure~\ref{fig:sys}(a)) spans $R_w(\Kperp) \in [1.139, 1.219]$ ($\Delta R_w = -0.042 / +0.039$) for $\BB[4,6]$ and $[1.139, 1.220]$ ($\Delta R_w = -0.042 / +0.038$) for $\DD[4,6]$. Even at the lower edge of this $-20\%$ initial-saturation-scale band ($R_w = 1.139$), the central-$\pPb$ saturation prediction remains well above the collinear \texttt{nNNPDF3.0} ($1.060$ for $\BB$, $1.097$ for $\DD$) and Gaussian-kick ($1.023$ and $1.034$) baselines. All other theoretical and instrumental variations are smaller by nearly an order of magnitude:
\begin{enumerate}
  \item \emph{Nonperturbative Sudakov parameter $\CNP$}: As shown in figure~\ref{fig:sys}(b), scanning $\CNP$ by more than a factor of two across $\CNP \in \{1.5, 1.9, 2.3, 2.9, 3.5\}$ shifts the excess broadening $R_w(\Kperp) - 1$ in $[4,6]\gev$ by less than $\pm 0.20\%$ for central $\BB$ ($\mathcal{F}_{\mathrm{NP}} = 93.4$), $\pm 0.10\%$ for minimum-bias $\BB$ ($\mathcal{F}_{\mathrm{NP}} = 132.8$), and $\pm 0.24\%$ for central $\DD$ ($\mathcal{F}_{\mathrm{NP}} = 72.2$).
  \item \emph{Sudakov--hard threshold matching scale $c_{\mathrm{match}}$}: Varying the matching parameter in eq.~\eqref{eq:W_match} by a factor of four across $c_{\mathrm{match}} \in [0.5, 2.0]$ changes $R_w(\Kperp)$ in $[4,6]\gev$ by at most $\Delta R_w = 0.0089$ ($R_w \in [1.185, 1.196]$) for $\BB$ and $\Delta R_w = 0.0041$ ($R_w \in [1.171, 1.177]$) for $\DD$.
  \item \emph{Perturbative dead-cone Sudakov mass term}: As demonstrated in section~\ref{sec:results_deadcone}, setting $m_{\mathrm{Sud}} \to 0$ inside $S_P(b; \mu, m_Q)$ shifts $R_w(\Kperp)$ by only $-0.0007$ in $\BB[4,6]$ and $+0.000002$ in $\DD[4,6]$.
  \item \emph{LHCb tracking and azimuthal resolution}: Applying independent Gaussian smearing of $\sigma(p_T)/p_T = 1.5\%$ and $\sigma(\phi) = 10\,\mathrm{mrad}$ to each reconstructed meson shifts the paired width ratio $R_w(\Kperp)$ by only $\Delta R_w = -0.0012$ for $\BB[4,6]$ and $-0.0019$ for $\DD[4,6]$ ($\Delta R_{Q_{50}} = +0.0004$ and $-0.0014$).
\end{enumerate}
In the lower $[2,4]\gev$ bin, where the flattened central-$\pPb$ $\Kperp$ plateau makes the parametric Gaussian fit $R_w$ more sensitive to $c_{\mathrm{match}}$ ($\Delta R_w = 0.054$ for $\BB[2,4]$ and $0.099$ for $\DD[2,4]$) and resolution smearing ($\Delta R_w = 0.0037$ and $0.050$), the non-parametric median quantile ratio $R_{Q_{50}}$ remains immune to detector resolution ($\Delta R_{Q_{50}} = +0.0003$ for $\BB[2,4]$ and $+0.0004$ for $\DD[2,4]$) and preserves a clean $Q_{s0,\mathrm{Pb}}^2 \pm 20\%$ hierarchy ($R_{Q_{50}} \in [1.034, 1.051]$ for $\BB[2,4]$ and $[1.042, 1.062]$ for $\DD[2,4]$). Combining $R_{Q_{50}}$ at $[2,4]\gev$ with $R_w(\Kperp)$ and $\mathcal{D}_{12}(\Kperp)$ at $[4,6]$ and $[6,8]\gev$ therefore provides a well-controlled experimental program across the full LHCb forward kinematic range.

\section{Summary and outlook}
\label{sec:conclusions}

In this paper, we have formulated and numerically evaluated a set of energy-weighted and single-axis-projected correlation observables—the heavy-flavor-tagged transverse energy--energy correlator $\SigmaHF^{(n)}(\tau)$ (HF-TEEC), its mirror-antisymmetric combination $A_{\mathrm{HF}}^{(n)}(\tau)$ (HF-ATEEC), the orthogonal pair-imbalance projection $\Kperp \equiv \aT$, and its one-dimensional Fourier-cosine transform $\mathcal{R}_{\cos}^{\mathrm{proj}}(b)$—for forward $\DD$ and $\BB$ production in proton-proton and proton-nucleus collisions. Our construction addresses the three main backgrounds that complicate the extraction of small-$x$ gluon saturation from tagged heavy-meson pairs at LHCb: longitudinal fragmentation recoil along the di-meson axis, nuclear-enhanced double parton scattering (DPS), and nonperturbative Sudakov uncertainties. Analytically, we showed that weighting the pair cross section by $(E_{T1}E_{T2})^n$ shifts the sampled heavy-quark fragmentation distribution toward larger momentum fractions $\xi$ via a positive cumulant covariance and reduces the relative DPS pedestal across wide $p_T$ intervals, while subtracting the mirrored near-side hemisphere in HF-ATEEC cancels any flat DPS background and all even azimuthal harmonics identically. In transverse-momentum space, projecting $\bq_T = \bp_1 + \bp_2$ onto the axis orthogonal to $\bP_h = \frac{1}{2}(\bp_1 - \bp_2)$ eliminates linear $\mathcal{O}(\xi_1 - \xi_2)$ longitudinal fragmentation recoil exactly, yielding a cusp-free distribution peaked at $\Kperp = 0$ whose one-dimensional Fourier-cosine transform cancels common Sudakov factors in the factorized back-to-back limit.

We implemented these observables within the massive hybrid Color Glass Condensate framework by combining HERA-constrained Balitsky running-coupling BK dipole amplitudes with threshold-matched SCET/bHQET massive Sudakov resummation and finite-mass KKKS08 ($c \to D^0$) and DGLAP-evolved KKSS ($b \to B^0$) fragmentation functions. Benchmarking the leading-order angular calculation against published LHCb forward $\DD$ data in $\pp$ ($\sqrt{s} = 7\tev$) and minimum-bias $\pPb$ ($\sqrt{s_{NN}} = 8.16\tev$) collisions demonstrated that the threshold-matched CGC+Sudakov formalism accurately describes the away-side acoplanarity slope and its nuclear broadening, whereas the narrow near-side charm peak and mid-angle $\pPb$ pedestal reflect NLO $g \to c\bar{c}$ splittings and nuclear-enhanced DPS ($F_{\mathrm{DPS}}^{p\mathrm{Pb}} \simeq 3\text{--}4$). Applying the mirror-antisymmetric HF-ATEEC directly to the symmetrically binned LHCb $\pPb$ data removes the flat DPS pedestal algebraically and exposes the away-side nuclear broadening across $\tau \in [0.10, 0.50]$. Across wide $p_T$ windows at $\sqrt{s_{NN}} = 8.16\tev$, linear ($n=1$) and quadratic ($n=2$) transverse-energy weighting reduces the relative DPS-to-SPS fraction by $15\%$ and $30\%$ in $\DD[2,8]\gev$ ($5.7\%$ and $11.5\%$ in $\BB[3,9]\gev$), complementing the exact reflection cancellation of HF-ATEEC in narrower bins.

At forward LHC Run~3--4 kinematics ($\sqrt{s_{NN}} = 8.16\tev$, $2.7 < y_{1,2} < 3.7$), our comparative analysis of the smooth $\Kperp \equiv \aT$ spectra against collinear nuclear shadowing (\texttt{nNNPDF3.0}) and linear Gaussian broadening established a clear quantitative hierarchy across transverse-momentum windows and heavy-quark flavors. In the lowest $[2,4]\gev$ bin, strong nuclear saturation in central $\pPb$ collisions flattens the back-to-back charm peak into a plateau over $\Kperp \lesssim 2\gev$, where the non-parametric median quantile ratio $R_{Q_{50}} = 1.060$ ($\DD$) and $1.047$ ($\BB$) provides a stable measure of broadening. In the intermediate $[4,6]\gev$ window, central (minimum-bias) CGC saturation predicts Gaussian width ratios of $R_w(\Kperp) = 1.188$ ($1.129$) for $\BB$ and $1.175$ ($1.121$) for $\DD$, exceeding both collinear \texttt{nNNPDF3.0} shadowing ($1.060$ and $1.097$) and linear Gaussian broadening ($1.023$ and $1.034$). The large excess of the full CGC calculation over the Gaussian-kick baseline—which shares the exact same saturation-scale shift $\Delta_{\mathrm{Pb}}^2(x_A) = Q_{s,\mathrm{Pb}}^2 - Q_{s,p}^2$—proves that heavy-flavor acoplanarity broadening at $p_T \sim 4\text{--}6\gev$ is governed by coherent scatterings in the power-law semi-hard tail ($l \sim 2\text{--}6\gev$) of the rcBK dipole rather than by soft Gaussian diffusion. Stepping up to $[6,8]\gev$ reduces the CGC broadening to $1.143$ ($\BB$) and $1.131$ ($\DD$) as $\Qs(x_A)/\mu$ decreases, whereas the collinear nPDF baseline exhibits a much flatter $p_T$ dependence; this distinct scale evolution is captured directly by the upward tilt of the two-scale double ratio $\mathcal{D}_{12}(\Kperp)$.

Through a controlled mechanism decomposition, we also clarified the physical origin of the heavy-quark mass hierarchy in self-normalized ratios. In the individual $\pp$ and $\pPb$ spectra, the SCET/bHQET dead-cone threshold freezes the final-state double logarithm below $\bar{\mu} = m_Q$ and lowers $S_P(b; \mu, m_b)$ once $m_Q > e^{1/2}\mu_b$ ($b \gtrsim 0.43\gev^{-1}$ for bottom, whereas charm sits right at the crossover $e^{1/2}\mu_b^{\min} \simeq 1.23\gev$). In self-normalized $\pPb/\pp$ ratios, however, a control calculation with $m_{\mathrm{Sud}} \to 0$ inside $S_P$ shifts $R_w(\Kperp)$ by less than $0.1\%$, demonstrating that the common perturbative Sudakov factor cancels to high accuracy between proton and nuclear targets. Consequently, the wider separation between CGC saturation and collinear shadowing in $\BB$ relative to $\DD$ stems from harder $b \to B^0$ fragmentation ($\langle\bar{\xi}\rangle \simeq 0.806$ vs.\ $0.664$), smaller probed $x_A$ at matched meson $p_T$, and weaker DGLAP scale variation of $R_g^{\mathrm{Pb}}(x_A, \mu)$ at the higher hard scale $\mu \sim 2m_{Tb}$ of bottom-pair production. In conjugate impact-parameter space, the finite-acceptance cosine ratio $\mathcal{R}_{\cos}^{\mathrm{proj}}(b)$ in $\BB[4,6]\gev$ tracks the underlying AAMQS dipole ratio $S_A(b)/S_p(b)$ closely for $b \lesssim 0.30\gev^{-1}$ ($0.945$ vs.\ $0.951$ in central $\pPb$ at $b_* = 0.30\gev^{-1}$), while larger impact parameters $b \gtrsim 0.35\gev^{-1}$ receive finite-acceptance truncation corrections.

Finally, our one-dimensional rcBK nuclear-opacity scan across $\nu_A = Q_{s0,A}^2/Q_{s0,p}^2 \in [1.0, 4.33]$ and systematic uncertainty audit confirmed that the saturation broadening signal scales monotonically from $\pp$ through $0\text{--}10\%$ central $\pO$ ($R_w = 1.028$ in $[4,6]\gev$) and minimum-bias $\pPb$ ($1.133$ for $\BB$, $1.112$ for $\DD$) to central $\pPb$ ($1.180$ and $1.181$). Varying the initial lead saturation scale $Q_{s0,\mathrm{Pb}}^2$ by $\pm 20\%$ yields $R_w(\Kperp) \in [1.139, 1.219]$ in $[4,6]\gev$, which represents the primary theoretical uncertainty yet remains well separated from both non-saturation baselines. All other systematic variations in $[4,6]\gev$—the five-point nonperturbative Sudakov scan $\CNP \in [1.5, 3.5]$ ($\mathcal{F}_{\mathrm{NP}} = 72\text{--}133$), the factor-of-four matching-scale variation $c_{\mathrm{match}} \in [0.5, 2.0]$ ($\Delta R_w \le 0.009$), and LHCb momentum and azimuthal resolution smearing ($\Delta R_w \le 0.002$)—are smaller by nearly an order of magnitude.

Several extensions of this work will further sharpen comparisons with upcoming LHC and future Electron-Ion Collider measurements. First, incorporating complete next-to-leading-order massive hybrid-CGC hard coefficients and collinearly stabilized NLO small-$x$ evolution~\cite{Shi:2021hwx,Wang:2022fwj,Boussarie:2025rot} will reduce perturbative scale uncertainties and fill in the near-side $g \to Q\bar{Q}$ collinear splitting region at $\tau \to 1$. Second, extending the target description beyond the large-$N_c$ Gaussian dipole factorization to include finite-$N_c$ quadrupole correlators and impact-parameter-dependent JIMWLK evolution will refine predictions in the deep-saturation regime at $p_T \in [2,4]\gev$. Third, on the experimental side, folding our prompt $\BB$ predictions with $B \to J/\psi + X$ and $B \to D + X$ decay kinematics and combining $\Kperp$ energy correlators with event-activity or intra-jet multiplicity selections~\cite{Duan:2026eec,Wang:2026mult} will enable multi-differential tests of nonlinear gluon dynamics in LHCb Run~3 and Run~4 $\pO$ and $\pPb$ collisions.

\acknowledgments
This work is supported in part by the National Natural Science Foundation of China (NSFC) under Grant No.~1234710148, and in part by the China Postdoctoral Science Foundation under Grant No.~2023M742098.

\appendix
\section{\texorpdfstring{Analytic $\Kperp$ inversion, Jacobian, and numerical checks}{Analytic Kperp inversion, Jacobian, and numerical checks}}
\label{app:kperp}

Finite-width histograms of $(\bp_1,\bp_2,\delta\phi)$ introduce statistical fluctuations into $\dd\Sigma/\dd\Kperp$ that can be amplified in nuclear and two-scale ratios $\mathcal{D}_{12}(\Kperp)$. Because $\Kperp$ in eq.~\eqref{eq:Kperp_def} is a strictly monotonic function of the away-side deflection angle $\delta\phi \in [0, \pi/2]$ at fixed meson transverse momenta $(p_{T1}, p_{T2})$, binning noise can be eliminated by inverting $\Kperp(p_{T1}, p_{T2}, \delta\phi)$ analytically.

Writing $c \equiv \cos\delta\phi \in [0, 1]$ on the away-side hemisphere and squaring eq.~\eqref{eq:Kperp_def} yields a quadratic equation in $c$:
\begin{equation}
  4 p_{T1}^2 p_{T2}^2 (1 - c^2) = \Kperp^2 \left(p_{T1}^2 + p_{T2}^2 + 2 p_{T1} p_{T2} c\right).
  \label{eq:kp_quad}
\end{equation}
For any physically accessible value $0 \le \Kperp < K_{\max}(p_{T1}, p_{T2}) \equiv 2 p_{T1} p_{T2} / \sqrt{p_{T1}^2 + p_{T2}^2}$ (reached at $\delta\phi = \pi/2$), eq.~\eqref{eq:kp_quad} admits a unique nonnegative root:
\begin{equation}
  c(\Kperp; p_{T1}, p_{T2})
  = \frac{-\Kperp^2 + \sqrt{\left(4 p_{T1}^2 - \Kperp^2\right)\left(4 p_{T2}^2 - \Kperp^2\right)}}{4 p_{T1} p_{T2}},
  \qquad \delta\phi = \arccos c.
  \label{eq:kp_inv}
\end{equation}
Implicit differentiation of eq.~\eqref{eq:kp_quad} gives the exact change-of-variables Jacobian between $\delta\phi$ and $\Kperp$,
\begin{equation}
  \left|\frac{\partial\delta\phi}{\partial\Kperp}\right|
  = \frac{\left(p_{T1}^2 + p_{T2}^2 + 2 p_{T1} p_{T2} c\right)^{3/2}}{2 p_{T1} p_{T2} \left[\left(p_{T1}^2 + p_{T2}^2\right) c + p_{T1} p_{T2}\left(1 + c^2\right)\right]},
  \label{eq:kp_def_jac}
\end{equation}
which remains strictly finite and smooth across the entire away-side interior, including the back-to-back endpoint $\Kperp \to 0$ ($c \to 1$) where $|\partial\delta\phi/\partial\Kperp|_{\Kperp=0} = (p_{T1} + p_{T2})/(2 p_{T1} p_{T2})$. Evaluating the 10-dimensional phase-space integral over $(p_{T1}, p_{T2}, y_1, y_2, \xi_1, \xi_2, \bl_1, \bl_2)$ at fixed $\Kperp$ knots using shared scrambled Sobol sequences removes histogram noise and keeps discrete second differences of the $\BB[4,6]\gev$ nuclear ratio below $10^{-3}$ across $0 \le \Kperp \le 3.6\gev$.

All analytical identities and numerical routines have been cross-checked independently against direct quadrature and finite differences. The finite-mass fragmentation map~\eqref{eq:frag_map}, $\tau$-Jacobian conservation~\eqref{eq:tau_def}, ATEEC reflection cancellation~\eqref{eq:thm2b}, exact $\Kperp \equiv \aT$ equivalence, and vanishing linear fragmentation derivative $|\partial\Kperp/\partial\epsilon|_{\epsilon=0}$ hold to machine precision ($<10^{-13}$). The Mellin hardening derivative~\eqref{eq:thm1}, dead-cone derivative~\eqref{eq:deadcone_deriv}, 2D-to-1D cosine reduction~\eqref{eq:1d_cosine}, and Jacobian~\eqref{eq:kp_def_jac} agree with numerical evaluation to $10^{-10}\text{--}10^{-7}$. Finally, the ideal factorized Sudakov cancellation~\eqref{eq:thm4} holds to $< 10^{-4}$, and the signed cubic-spline Hankel transform preserves $\int\dd^2\bl\,F_A = 1$ to better than $5\times 10^{-4}$ across all proton, oxygen, and lead dipole tables.

\section*{Data and code availability}
The analysis scripts and generated numerical arrays are available from the author upon reasonable request.

\bibliographystyle{JHEP}
\bibliography{references}

\end{document}